\documentclass[fleqn,usenatbib]{mnras}

\usepackage{newtxtext,newtxmath}

\usepackage[T1]{fontenc}
\usepackage{threeparttable}
\usepackage{multirow}
\usepackage{ulem}

\DeclareRobustCommand{\VAN}[3]{#2}
\let\VANthebibliography\thebibliography
\def\thebibliography{\DeclareRobustCommand{\VAN}[3]{##3}\VANthebibliography}

\usepackage{graphicx}	
\usepackage{amsmath}	
\usepackage{bm}   

\title[SOAR/GHTS transmission spectroscopy of WASP-69 b]{Tentative detection of titanium oxide in the atmosphere of WASP-69\,b with a 4m ground-based telescope}

\author[Qinglin Ouyang et al.]{
Qinglin Ouyang$^{1,3}$,
Wei Wang$^{2}$\thanks{E-mail: wangw@nao.cas.cn},
Meng Zhai$^{1,2}$\thanks{E-mail: mzhai@nao.cas.cn},
Guo Chen$^{4}$,
Patricio Rojo$^{5}$,
Yujuan Liu$^{2}$,
Fei Zhao$^{2}$
\newauthor{
Jia-Sheng Huang$^{1}$,
and Gang Zhao$^{2,3}$}
\\
$^{1}$Chinese Academy of Sciences South America Center for Astronomy (CASSACA), National Astronomical Observatories, Chinese Academy of Sciences, \\Datun Road A20, Beijing 100101, China\\
$^{2}$CAS Key Laboratory of Optical Astronomy, National Astronomical Observatories, Chinese Academy of Sciences, Datun Road A20, Beijing 100101, China\\
$^{3}$School of Astronomy and Space Science, University of Chinese Academy of Sciences (UCAS), Yuquan Road A19, Beijing 100049, China\\
$^{4}$CAS Key Laboratory of Planetary Sciences, Purple Mountain Observatory, Chinese Academy of Sciences, Nanjing 210023, People’s Republic of China\\
$^{5}$Departamento de Astronomía, Universidad de Chile, Camino El Observatorio 1515, Las Condes, Santiago, Chile
}

\date{Accepted XXX. Received YYY; in original form ZZZ}

\pubyear{2022}

\begin{document}
\label{firstpage}
\pagerange{\pageref{firstpage}--\pageref{lastpage}}
\maketitle

\begin{abstract}
Transiting planets provide unique opportunities for the atmospheric characterization of exoplanets as they can reveal composition and the temperature structures at the day-night terminator regions in planetary atmospheres, and help understand the atmospheric process and formation environments of exoplanets. Here, we present the optical transmission spectroscopic study of an inflated Saturn-mass planet WASP-69\,b, obtained by the 4-meter ground-based telescope Southern Astrophysical Research Telescope (SOAR). We obtain spectroscopic transit light curves in 20 passbands from 502 to 890 nm, and fit them using Gaussian Processes and an analytical transit model to obtain independent transit depths for each. The derived transmission spectrum of WASP-69\,b shows a slope with absorption depth increasing towards blue wavelengths, indicating a Rayleigh scattering in the atmosphere consistent with previous works. The retrieval analysis yields a tentative detection of TiO absorption feature in the transmission spectrum. We present the first results from the SOAR telescope to characterize exoplanetary atmospheres proving its capability and precision for hot Jupiters around bright stars in an area dominated by results from large ground-based telescopes or space telescopes.
\end{abstract}

\begin{keywords}
methods: data analysis -- techniques: spectroscopic -- stars: individual (WASP-69) -- planetary systems -- planets and satellites: atmospheres.
\end{keywords}



\section{Introduction}
\label{sec:Introduction}

Thousands of exoplanets have been found since the first exoplanet orbiting a Sun-like star, 51 Pegasi b, was discovered by \citet{MayorQueloze1995}. Frontier research has been rapidly moving towards the detailed characterization of exoplanet atmospheres. Transits are one of the most important and widely used methods in the study of exoplanets, via which one can detect and identify small and periodic dimming in photometric light curves as the planet blocks a certain amount of light from its host star as it transits. Furthermore, during a transit one can also obtain a transmission spectrum (i.e., different transit depths at different wavelengths) by multi-band photometric or spectroscopic transit observations, which can be used to infer the chemical compositions of planetary atmospheres at day-night terminator \citep[e.g., ][]{Charbonneau2002, Deming2013, Sing2016, Nikolov2016, Parviainen2018} to constrain the existence of clouds/hazes in the atmosphere \citep[e.g., ][]{Gibson2013_cloud, Kreidberg2014, Wakeford2017}, and/or to estimate atmospheric mass loss \citep[e.g., ][]{Vidal-Madjar2003, Lecavelier2010, Ehrenreich2015, Spake2018}.

Low-density hot Jupiters (HJs) are the most preferred targets for atmospheric studies with transmission spectrum because of their relatively strong spectral signals. These planets are close and usually tidally-locked to their parent stars. Most have been heated to more than 1,000\,K and inflated, usually resulting in large scale heights ($H$) and thus large spectral signals. Until now, many species have been detected in the atmosphere of HJs, e.g., Na~\citep{Redfield2008, Nikolov2016, Casasayas-Barris2017, Nikolov2018}, K~\citep{Sing2015, Chen2018}, TiO~(\citealt{Haynes2015,Nugroho2017,Sedaghati2017,Sedaghati2021}; but cf. \citealt{Espinoza2019}), VO~\citep{Evans2017,Evans2018}, He~\citep{Nortmann2018, Spake2018}, H$_2$O~\citep{McCullough2014, Evans2017}, and CO~\citep{Sheppard2017}. 

WASP-69\,b is a warm and highly inflated Saturn-mass planet ($M_\textnormal{p} = 0.260 \pm 0.017~M_\textnormal{Jup}$, $R_\textnormal{p}=1.057 \pm 0.047~R_{\textup{Jup}}$, $T_\textnormal{eq}=963 \pm 18$ \,K), orbiting a K-type host star with a period of $\sim$3.868 days~\citep{Anderson2014}. It has a large scale height with $H \sim 635$\,km, which amounts to $\sim 0.0011$ of stellar radius, and each scale height of atmosphere contributes $\sim 295$\,ppm to the transit depth for WASP-69\,b. Due to its low density and inflated atmosphere, WASP-69\,b is one of the most promising and interesting targets for transmission spectroscopy, and quite a number of works have been carried out since its discovery.

For atomic species, \citet{Casasayas-Barris2017} detected Na using the high-resolution spectra obtained with the HARPS-North spectrograph \citep{Mayor2003, Cosentino2012} at the 3.56\,m Telescopio Nazionale Galileo, which was then confirmed by a similar high-resolution study by \citet{Khalafinejad2021} using CARMENES \citep{Quirrenbach2018}. \citet{Nortmann2018} detected neutral helium by the He{\sc i} triplet 10830\,\AA\ in NIR, and proposed that helium is escaping from WASP-69\,b, trailing behind the planet. The helium escape was then confirmed by a more recent study based on ultra narrow band photometric observation~\citep{Vissapragada2020}.

Using their HST/WFC3 low-R NIR transmission spectra, \citet{Tsiaras2018} detected the water feature at 1.4 $\mu$m, and found an overlying slope of a larger planet radius towards blue wavelengths. Later on, \citet{Murgas2020} obtained an optical transmission spectrum of WASP-69\,b with GTC/OSIRIS, and confirmed the slope increase towards blue wavelengths, which was suggested to be caused by Rayleigh scattering. However, they could not rule out activity as causing the slope they saw. On the other hand, \citet{Khalafinejad2021} performed a joint analysis on the HST/WFC3 and GTC/OSIRIS data, and claimed strong Rayleigh scattering, solar to super-solar water abundance, and a highly muted Na feature. \citet{Estrela2021} independently analysed the archival HST/STIS and HST/WFC3 data, and concluded that aerosol hazes should exist in the atmosphere of WASP-69\,b. 

In this paper, we present a transmission spectroscopic study of WASP-69\,b, using two transit observations obtained with the Goodman High Throughput Spectrograph \citep[GHTS, ][]{Goodman_intro2004} installed on the 4.1 m Southern Astrophysical Research Telescope (SOAR), in Cerro Pachón, Chile. Our goal is to search for species with strong absorption features in the optical passband, and to confirm or reject the presence of clouds/hazes in the atmosphere of WASP-69\,b. While most of such studies were carried out using the space-borne telescopes like HST, or ground-based telescopes with aperture size larger than 5.1\,m, attempts have been made in recent years by using the much less expensive 4m class ground-based telescopes, including the Low Resolution Ground-Based Exoplanet Survey using Transmission Spectroscopy (LRG-BEASTS) using the William Herschel Telescope and New Technology Telescope~\citep{Kirk2016, Kirk2017, Kirk2018, Kirk2019, Alderson2020, Kirk2021, Ahrer2022}, and this study, all of which are proved to be successful.

This manuscript is organized as follows. In Sect.~\ref{sec:Observation and data reduction} we describe the  observations and methods used in this work. In Sect.~\ref{sec:Light curve analysis}, we present our light curve analysis. In Sect.~\ref{sec:Results} we describe the optical transmission spectrum obtained for WASP-69\,b and the atmosphere model that we have considered, following by discussion in Sect.~\ref{sec:Discussion}. Sect.~\ref{sec:Conclusions} summarizes our conclusions of this work.

\section{Observations and data reduction}
\label{sec:Observation and data reduction}
\subsection{Observations}
\label{sec:observations} 
\begin{figure}
	\includegraphics[width=\columnwidth]{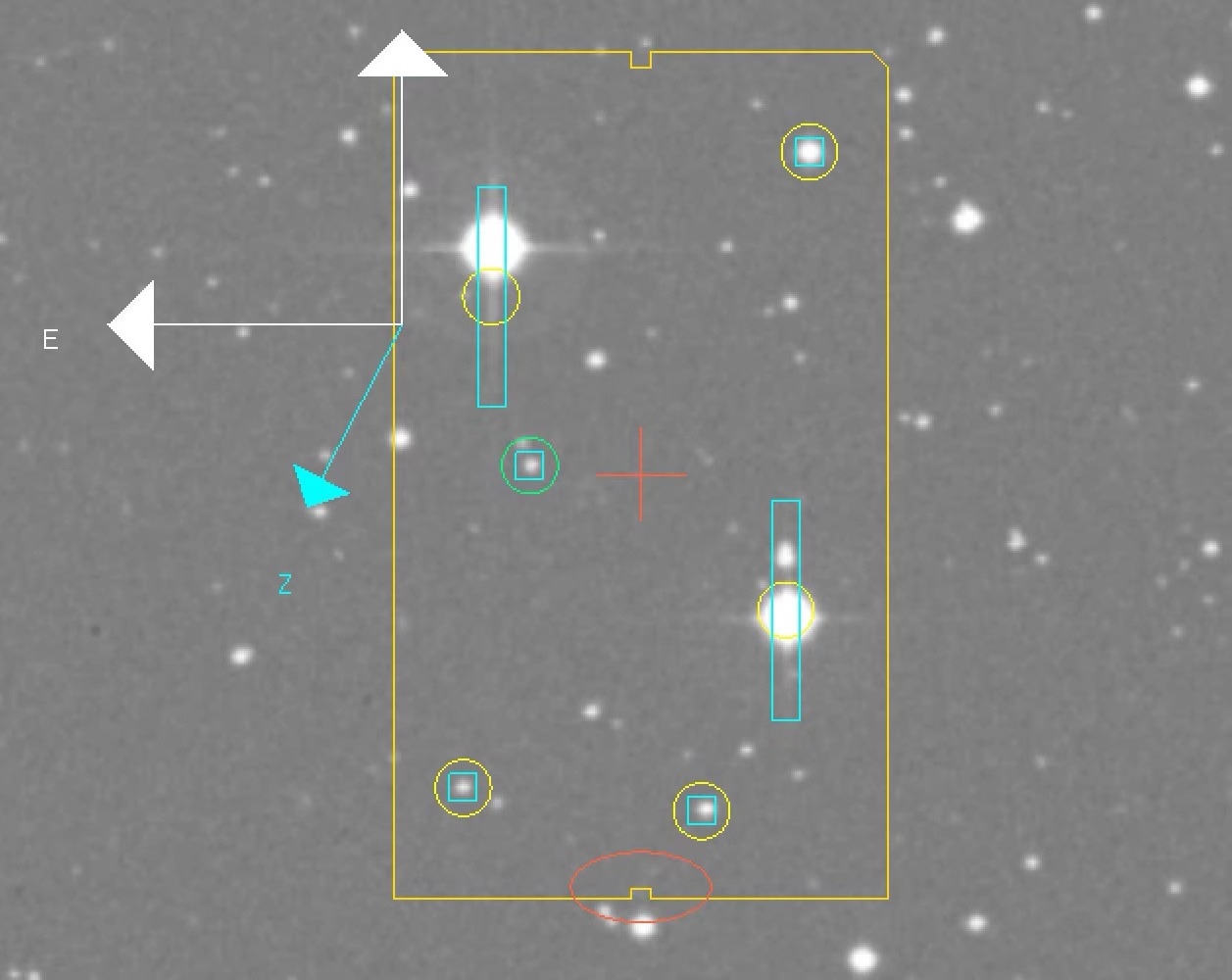}
    \caption{The field of view (FOV) overplotted as the yellow rectangle on the 2MASS sky image. The two long wide-slits for WASP-69 and the reference star TYC 5187-1718-1 are shown as the top left cyan rectangles and the lower right one, respectively. The small cyan squares with yellow circles are for the guiding stars, while that with a cyan circle is a short wide-slit. The center of each circle represents the center of each box or square.}
    \label{fig:WASP69_FOV}
\end{figure}

Two transits of WASP-69\,b were observed using GHTS at SOAR (Program CN2017A-31, PI: Wei Wang). We used GHTS in the Multi-Object Spectroscopy (MOS) mode with the field of view (FOV) of 3$\times$5 arcmin, in order to simultaneously observe the target and the reference star for the purpose of differential spectrophotometry. We chose the star TYC\,5187-1718-1 as the reference star, which is 171.82\arcsec\, away from the WASP-69 and has a brightness similar to the target star. The $V$-magnitudes for the target and reference stars are 9.87 and 10.60~\citep{Hog2000}, respectively. It is a pity that this is only one appropriate reference stars in the FOV. Two long wide-slits were carved in the slit mask that we designed for this observation, with one slit for WASP-69 and the other for the reference star. (cf. Figure~\ref{fig:WASP69_FOV}).

The slits were carved to have a projected width of 20\arcsec, to avoid slit loss that may occur in case of large seeing variation. The slits are  $\sim100$\arcsec long to allow enough room for sky background subtraction. The 400 M2 grating was used, which provides wavelength coverage 500-905\,nm and spectral resolution R$\sim$1850 with seeing size of $\sim0.45\arcsec$. The GG455 filter was used to block second-order contamination, and the 2 $\times$ 2 binning mode (0.30$\arcsec$ per binned pixel) was used for readout. 

The two observations were taken on the nights of July 15, 2017 (hereafter Night 1) and July 19, 2017 (hereafter Night 2), covering a UT window of 05:59-10:06 and 02:45-06:23, respectively. In total, we obtained 287 and 290 spectra for each night respectively, with exposure time of $30-50$ seconds. The detector readout time is 6 seconds, resulting a total duty cycle of 81\%. However, it was not photometric with thin cirrus in Night 1 and CCD movement during the observation, resulting in a flux jump in first several data points and a slightly larger point-to-point scatter in the transit light curve (Fig.~\ref{fig:Figure 2}). Therefore, for Night 1, first 23 and last 74 exposures with large scatter are excluded, while the rest 190 spectra are employed for the following light curve analysis. The details of both observations are shown in Table~\ref{tab:observaition details}.
\begin{table*}
	\centering
	\caption{Details of the two transit observations in this work.}
	\label{tab:observaition details}
	\begin{threeparttable}
	\begin{tabular}{lcccccc} 
		\hline
		Observation night & Tel./Instrument & Start time(UTC) & End time(UTC) & Exposure time(s) & Exposure numbers & Airmass range \\
		\hline
		2017-07-15 & SOAR/GHTS & 05:59:12 & 10:06:06 & 40/50 & 287 (190 used) & 1.11 $-$ 2.05 (1.11 $-$ 1.44)\tnote{a} \\
		2017-07-19 & SOAR/GHTS & 02:45:21 & 06:23:00 & 30/35/40 & 290 & 1.11 $-$ 1.61\\
		\hline
	\end{tabular}
    \begin{tablenotes}
        \item[a] The airmass range in the parentheses is for the 190 used exposures.
    \end{tablenotes}
    \end{threeparttable}
\end{table*}

\subsection{Data reduction}
\label{sec:Data reduction} 
The raw data were reduced mainly used the Python scripts developed by us. The 2D spectra were firstly corrected for overscan, bias and the flat field. The \texttt{dcr} package~\citep{Pych2012} was used to remove cosmic ray hits in 2D spectra images. The algorithm is based on a simple analysis of the histogram of the image data and reject those pixels with counts exceeding a certain value. We chose conservative \texttt{dcr} parameters so that the package won't mistake stellar signals for cosmic ray hits. As a result, there are still some cosmic ray hits remained close to the stellar spectra after applying \texttt{dcr}, which were then removed manually. For the 1D spectra extraction, we utilized the optimal extraction algorithm \citep{Horne1986}, and tested a series of aperture width from 17 pixels to 35 pixels to minimize the point-to-point scatter of the out-of-transit data. Finally, the aperture size of 34 pixels ($\sim 10.2\arcsec$) for Night 1 and 31 pixels ($\sim 9.3\arcsec$) for Night 2 were chosen, respectively. The initial wavelength solution was achieved by the HgArNe arc lamp spectra taken before and after science frames, using \texttt{identify} in IRAF package. This was later refined by aligning all the spectra in the wavelength domain, using the strong absorption lines, e.g., Na doublet, H$\alpha$ and telluric O$_2$. We used a second-order polynomial to fit the movements of all three absorption lines in each exposure relative to the first exposure. The time stamp of each spectra was centered on mid-exposure and converted into Barycentric Julian Dates in Barycentric Dynamical Time (BJD$_{\textup{TDB}}$; \citealt{Eastman2010}).

\begin{figure}
	\includegraphics[width=\columnwidth]{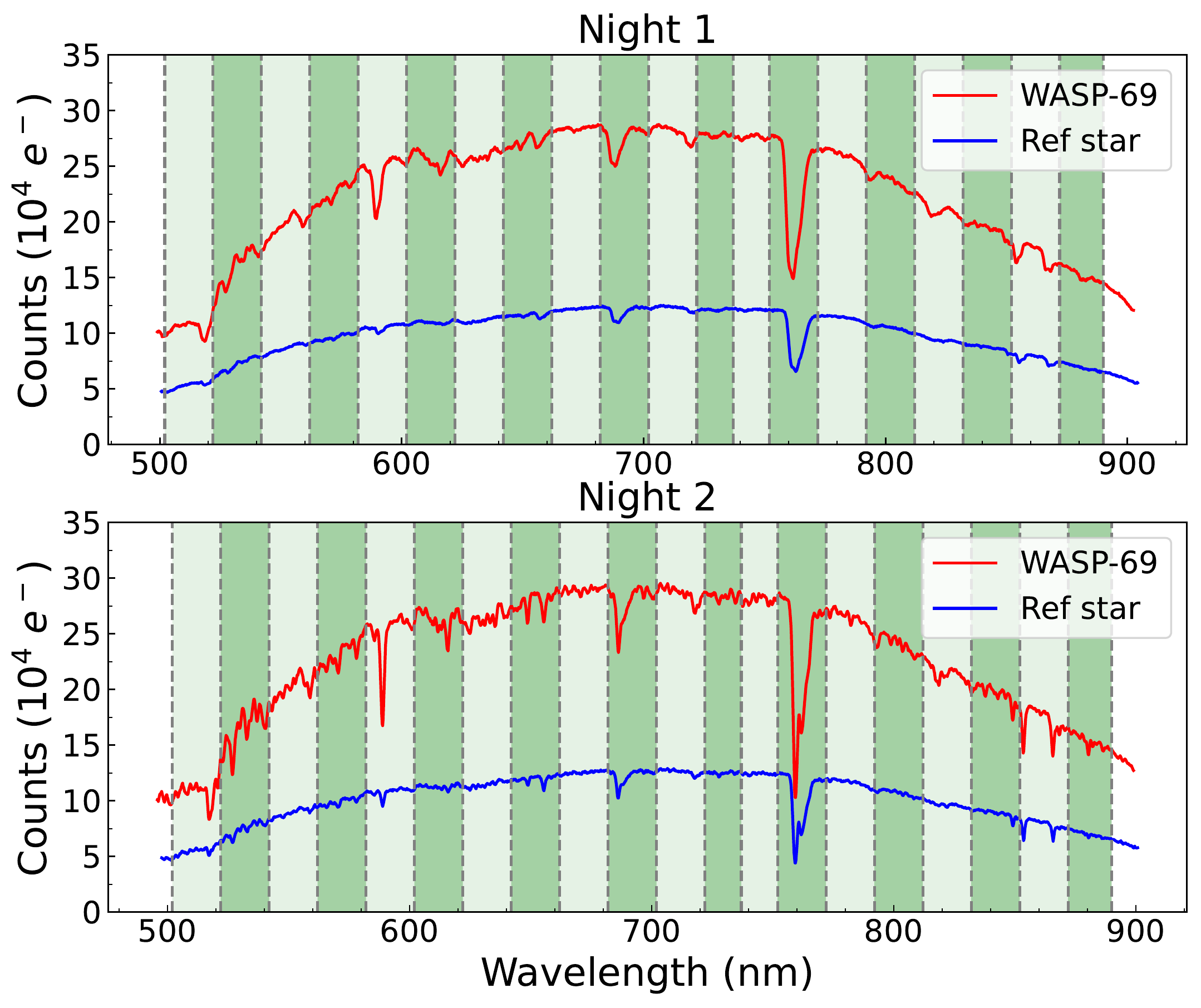}
    \caption{Example stellar spectra of WASP-69 (red) and reference star (blue) for two observing nights (top and bottom). The dark and light green shaded zones indicate the individual passbands used in this work.}
    \label{fig:Figure 1}
\end{figure}

\begin{figure*}
	\includegraphics[width=\textwidth]{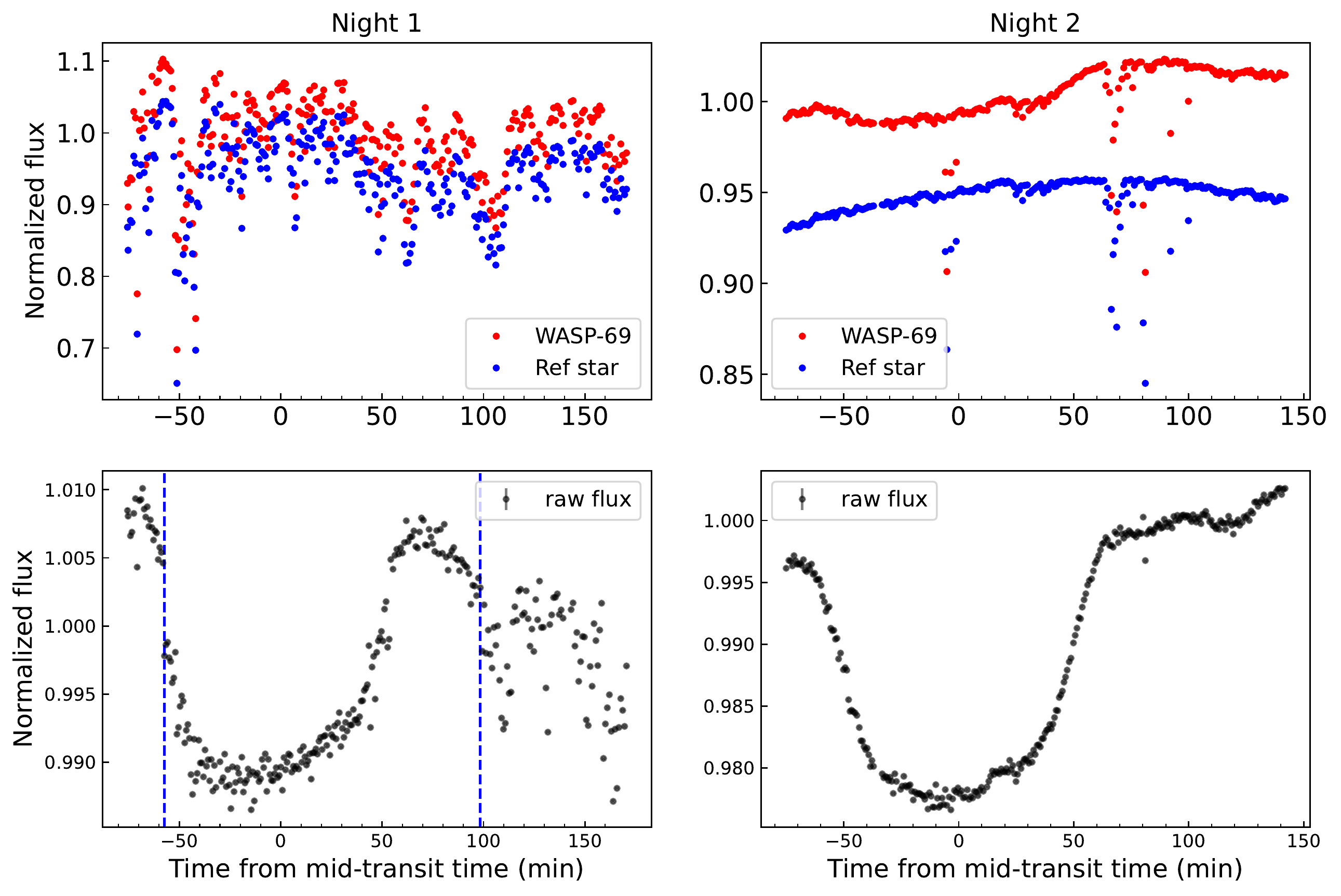}
    \caption{The white light curves of the two transits in Night 1 and Night 2. \textit{Top}: raw stellar light curves of WASP-69 (red) and reference star (blue) with the latter vertically shifted for clarity. \textit{Bottom}: the relative (target/reference) light curves. For Night 1, the data points  before $\sim-$60 min (as marked by the left blue dashed line) and after $\sim$95 min (the right blue dashed lines) are excluded from data analysis due to a sudden flux jump and a large point-to-point scatter, respectively.}
    \label{fig:Figure 2}
\end{figure*}

White light curves were created by summing the flux within the wavelength range from 502\,nm to 890\,nm for both the target and reference star, and then divided target light curve by reference light curve to remove telluric and instrument systematics. To create spectroscopic light curves, each spectrum was divided into 20 channels, with 2 channels of 15\,nm (722-737\,nm, 737-752\,nm) bin, 1 channel of 18\,nm (872-890\,nm) bin and 17 channels of 20\,nm bin. The reason of this passband division is to make sure the edge of each passband fall on the continuum rather than prominent stellar absorption lines. Fig.~\ref{fig:Figure 1} shows example spectra of the target and reference star for Night 1 \& 2, along with the 20 spectral channels shown in the shaded light and dark green. The raw light curves of both nights are shown in Figure.~\ref{fig:Figure 2}. As aforementioned, the Night 1 transit light curve has a large point-to-point scatter, due to non-perfect weather condition during the observation.

\section{Light curve analysis}
\label{sec:Light curve analysis}
In this section, we describe in detail how the obtained white and spectroscopic light curves were modeled, and how the transit depths were derived. In general, observed light curves consist of two independent components. One component represents the astrophysical signals, which can be described by analytic transit model such as the most widely-used \cite{Mandel2002} model (MA02). The other component is noise, including Gaussian noise (e.g., shot noise) which is uncorrelated with any parameters, and correlated noise of telluric or instrumental origins. The latter could be modelled in an analytic parametric form or by non-parametric Gaussian processes (GP; \citealt{Rasmussen2006}). The GP methodology was pioneered by \citet{Gibson2012(1), Gibson2012(2)} in the study of exoplanet atmospheres. It has the advantage of finding internal correlations between various parameters without providing specific functions and assessing how reliable these correlations are. It is now (one of) the most popular and widely-used technique in the study of transmission spectroscopy of exoplanets \citep[e.g., ][]{Gibson2013, Evans2015, Evans2017, Chen2021(1)}. In this work, we also use the GP method for the modelling of our light curves.

In practice, we first fitted the white light curves of both visits, with the transit described by the MA02 model using the Python package \texttt{batman}~\citep{Kreidberg2015}, while the correlated noise handled by GP via the Python package \texttt{george} \citep{Ambikasaran2015}. Then the 20 spectroscopic light curves were modelled using the transit parameters from the white light curves, to obtain wavelength-dependent transit depths. The procedures of our light curve analysis will be presented in details below. 

\subsection{White light curve}
\label{sec:White light curve} 
The white light curves of Night 1 and Night 2 were fitted separately. Using the GP method, we modelled each transit light curve as a GP:
\begin{equation}
    f(t, \bm{x}) \sim \mathcal{GP}(T(t,\bm{\phi}), \bm{\Sigma}(\bm{x}, \bm{\theta}))
	\label{eq:GP}
\end{equation}
where $T(t,\bm{\phi})$ is the transit function described by time $\textit{t}$ and transit parameters $\bm{\phi}$, expressed by the MA02 model via \texttt{batman} \citep{Kreidberg2015}. $\bm{\Sigma}$ is the covariance matrix, which is a function of additional parameters $\bm{x}$ and hyperparameters $\bm{\theta}$ of GP. Additional parameters vary with time during observations, such as airmass, seeing, spectral trace movement in spectral and spatial direction, etc. The element of covariance matrix is given by:
\begin{equation}
    \bm{\Sigma}(x_\textnormal{m}, x_\textnormal{n}) = k(x_\textnormal{m}, x_\textnormal{n})
\end{equation}
where $k$ is the kernel function (or covariance function), which decides each element in covariance matrix from additional parameters. We chose the Mat\'{e}rn $\nu$ = 3/2\, kernel in our modeling procedure, following the experiment of ~\citet{Gibson2013}. This kernel is given by:
\begin{equation}
    k(x_\textnormal{m}, x_\textnormal{n}) = A(1+\sqrt{3}R_\textnormal{mn})\,\textnormal{e}^{-\sqrt{3}R_\textnormal{mn}}
\end{equation}
where $A$ and $R$ are the hyperparameters specifying the amplitude and the scale of the kernel, respectively.

We implemented GP regression via the Python package \texttt{george}~\citep{Ambikasaran2015}, where the transit model was the mean function of the GP. For the selection of additional parameters for the GP, we explored all the combinations of the following parameters: time, airmass, seeing, and the position of target star in the spatial ($x$) and dispersion direction ($y$) and located the best parameter combination with the smallest Bayesian Information Criterion (BIC, \citealt{Schwarz1978}). We chose only time \textit{t} for Night 1 and the combination of time \textit{t} and seeing for Night 2 as the GP inputs after comparing the Bayesian Information Criterion values obtained for the visited combinations of parameters (cf. Table.~\ref{tab:BIC} for more details).

The quadratic limb darkening (LD) coefficients $u_{1}$ and $u_{2}$ were calculated for the wavelength range of 502-890\,nm using the Python package \texttt{PyLDTk}~\citep{Parviainen2015}, employing the stellar spectrum model library from \citet{Husser2013}, adopting the stellar effective temperature $T_\textnormal{eff}=4715\pm50$\,K, surface gravity log\,$g_{\star}=4.535\pm0.023$, and metallicity $Z=0.144\pm0.077$~\citep{Anderson2014} as the prior values.

We noticed that there is a sudden flux jump between the first 23 data points and other data points in Night 1 (Fig.~\ref{fig:Figure 2}). This may be due to a slight refinement of slit position after the first 23 spectra were taken, to place both target and reference stars well in slits to best avoid slit loss. Unfortunately, as the first 23 data points covered the before-transit baseline and a part of ingress, making it difficult to obtain a reliable assessment of this flux jump and thus they were excluded from light curve fitting. As a result, the useful before-transit baseline in Night 1 is missing, which hinders accurate determinations of transit parameters. Therefore, we decided not using the Night 1 data for atmospheric characterization of WASP-69\,b.

We used the Markov Chain Monte Carlo (MCMC) method via the Python package \texttt{emcee} \citep{Foreman-Mackey2013} to derive the marginalized posterior distribution of the investigated parameters, including the mid-transit time $T_\textnormal{mid}$, the planet-to-star radius ratio $R_\textnormal{p}/R_{\star}$, the scaled semi major axis $a/R_{\star}$, the orbit inclination $i$, $u_{1}$ and $u_{2}$, the hyperparameters $A$ and $R$, the white noise jitter $\sigma_\textnormal{w}$ which is to account for additional light-curve uncertainties not related to any parameter of investigation. The priors of all the investigated parameters are shown in Table~\ref{tab:Table 2}. The values of period $P$ and  eccentricity $e$ are fixed as 3.8681382 and 0, respectively~ \citep{Anderson2014}.

\begin{figure*}
	\includegraphics[width=\textwidth]{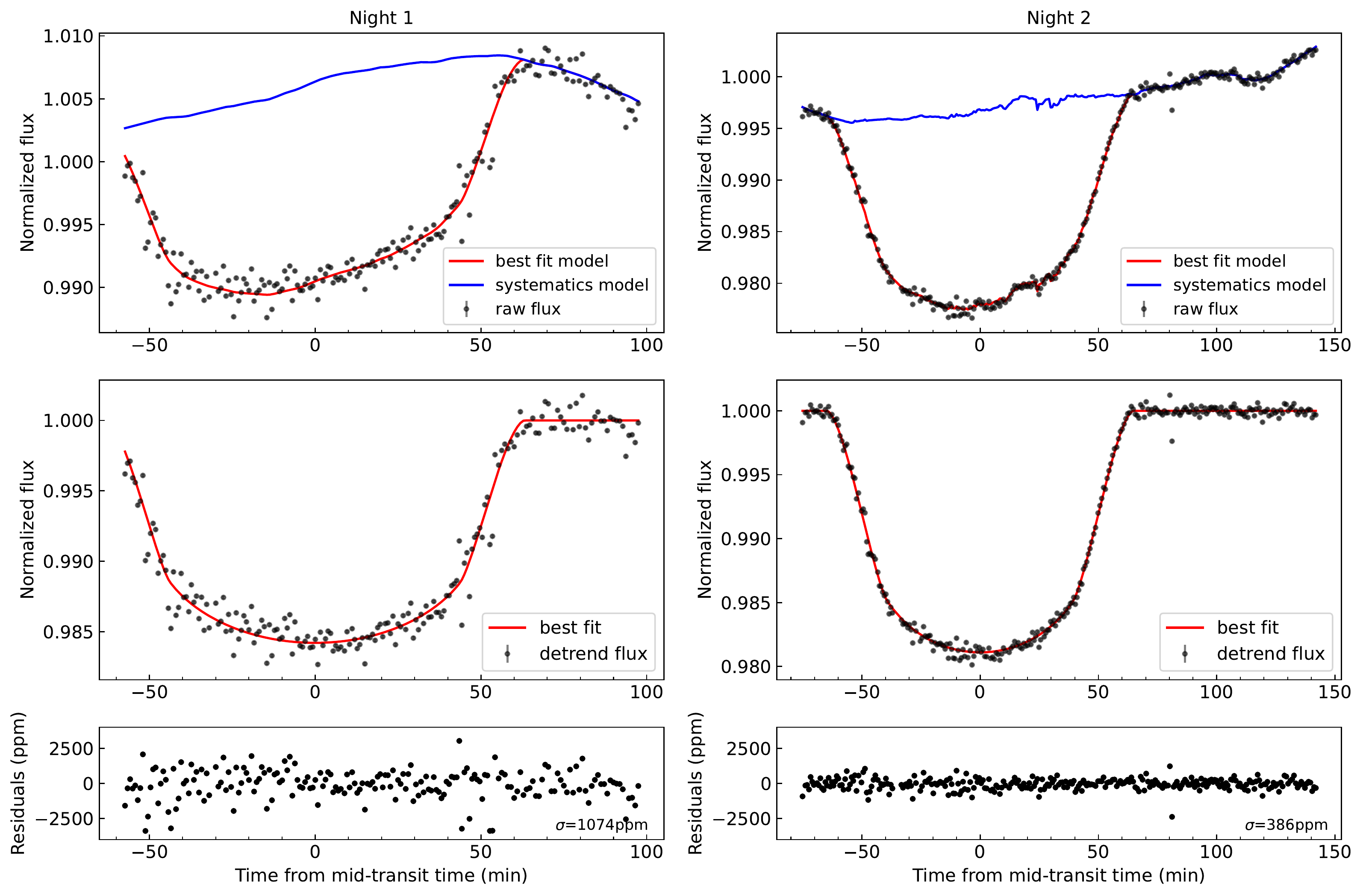}
    \caption{The raw (\textit{Top}), detrended (\textit{Medium}) light curves and residuals (\textit{Bottom}) of the Night 1 \& 2. The black dots, red and blue lines represent the data, the best fit models and the systematics models, respectively. The bottom panels show the residuals obtained by subtracting the best fit model from the data, with standard deviation of 1074\,ppm and 386\,ppm in Night 1 \& 2, respectively.}
    \label{fig:Figure 3}
\end{figure*}

The MCMC procedure consisted of 100 walkers, each one with 15000 steps, and the first 3000 steps were set as "burn-in" phase. The length of all chains were set to be more than 50 times autocorrelation time to assure convergence. The best posterior distribution of the transit parameters are given in Table~\ref{tab:Table 2}, and the best-fit models of the white light curves are given in Figure~\ref{fig:Figure 3}. As seen from Table~\ref{tab:Table 2}, the derived parameters from the two observations are consistent other than $R_\textnormal{p}/R_{\star}$ which is just greater than 1 sigma discrepant. The $R_\textnormal{p}/R_{\star}$ with the mean values and uncertainties are $0.11843_{-0.00794}^{+0.00831}$ (Night 1) and $0.13280_{-0.00574}^{+0.00551}$ (Night 2), respectively. This difference is reasonable given the relatively short out-of-transit baseline and large systematics of the Night 1.

\begin{table}
	\centering
	\caption{BIC calculated using the GP likelihood of different parameter combination as GP inputs in Night 1 \& 2. The best parameter sets are shown in bold.}
	\label{tab:BIC}
	\begin{threeparttable}
	\begin{tabular}{ccc} 
		\hline
		GP inputs & BIC of Night 1 & $\Delta$ BIC \\
		\hline
		time, airmass, $x$, $y$ & -2654.76 & 14.07 \\
		time, airmass, $x$ & -2659.76 & 9.07 \\
		time, airmass, $y$ & -2666.41 & 2.42 \\
		time, $x$, $y$ & -2660.00 & 8.83 \\
		time, $x$ & -2661.17 & 7.66 \\
		time, $y$ & -2662.22 & 6.61 \\
		$\textbf{time}$ & $\textbf{-2668.83}$  & \textbf{0}\\
		\hline
	    GP inputs & BIC of Night 2 & $\Delta$ BIC\\
		\hline
		time, airmass, seeing, $x$, $y$ & -4225.01 & 16.75 \\
		time, seeing, $x$, $y$ & -4226.70 & 15.06 \\
		time, seeing, $x$ & -4233.28 & 8.48 \\
		time, seeing, $y$ & -4238.57 & 3.19 \\
		time, $x$, $y$ & -4223.00 & 18.76 \\
		$\textbf{time, seeing}$ & $\textbf{-4241.76}$ & \textbf{0} \\
		time & -4236.30 & 5.46 \\
		\hline
	\end{tabular}
    \end{threeparttable}
\end{table}

\begin{table}
\renewcommand\arraystretch{1.5}
	\centering
	\caption{Parameters derived from the white light curve fitting.}
	\label{tab:Table 2}
	\begin{threeparttable}
	\begin{tabular}{lcc} 
		\hline
		\hline
		Parameters & Prior$^{a}$ & Posterior of Night 1\\
		\hline
		Period [days] & $3.8681382^{b}$  & ---\\
		\textit{e} & $0^{b}$ & ---\\	
		\noalign{\smallskip}
		\hline
		\noalign{\smallskip}
		& 2017-07-15 (Night 1) & \\
		$T_\textnormal{mid}$ [$\textnormal{MJD}^{c}$] & $\mathcal{U}(7949.75, 7949.83)$ & $7949.80802_{-0.00078}^{+0.00065}$\\
		$R_\textnormal{p}/R_{\star}$ & $\mathcal{U}(0.10, 0.20)$ & $0.11843_{-0.00794}^{+0.00831}$\\
		$a/R_{\star}$ & $\mathcal{U}(11.0, 15.0)$ & $13.50_{-0.81}^{+0.79}$\\
		$i$[deg] & $\mathcal{U}(80.0, 90.0)$ & $87.56_{-0.40}^{+0.39}$\\
		$u_{1}$ & $\mathcal{N}(0.612, 0.064)$ & $0.61155_{-0.06320}^{+0.06404}$\\
		$u_{2}$ & $\mathcal{N}(0.069, 0.086)$ & $0.06915_{-0.08736}^{+0.08630}$\\
        ln\,$A$ & $\mathcal{U}(-20, -1)$ & $-9.88_{-1.16}^{+1.95}$ \\
        ln\,$R_{t}$ & $\mathcal{U}(-15, 15)$ & $-4.39_{-1.06}^{+1.50}$ \\
		ln\,$\sigma_\textnormal{w}$ & $\mathcal{U}(-20, -1)$ & $-13.73_{-0.10}^{+0.11}$\\
		
		\noalign{\smallskip}
		\hline
		\noalign{\smallskip}
		            & 2017-07-19 (Night 2) & \\
		$T_\textnormal{mid}$ [$\textnormal{MJD}^{c}$] & $\mathcal{U}(7953.65, 7953.70)$ & $7953.67319_{-0.00036}^{+0.00035}$\\
	    $R_\textnormal{p}/R_{\star}$ & $\mathcal{U}(0.10, 0.20)$ & $0.13280_{-0.00574}^{+0.00551}$\\
		$a/R_{\star}$ & $\mathcal{U}(11.0, 15.0)$ & $12.68_{-0.27}^{+0.27}$\\
		$i$\,[$\deg$] & $\mathcal{U}(80.0, 90.0)$ & $87.03_{-0.16}^{+0.15}$\\
		$u_{1}$ & $\mathcal{N}(0.612, 0.063)$ & $0.61100_{-0.06527}^{+0.06561}$\\
		$u_{2}$ & $\mathcal{N}(0.069, 0.088)$ & $0.07352_{-0.08881}^{+0.09019}$\\
        ln\,$A$ & $\mathcal{U}(-20, -1)$ & $-12.30_{-0.82}^{+1.14}$ \\
        ln\,$R_{t}$ & $\mathcal{U}(-15, 15)$ & $-6.31_{-0.73}^{+0.90}$ \\
        ln\,$R_{seeing}$ & $\mathcal{U}(-15, 15)$ & $7.46_{-1.58}^{+1.89}$ \\
		ln\,$\sigma_\textnormal{w}$ & $\mathcal{U}(-20, -1)$ & $-15.77_{-0.10}^{+0.10}$\\
		\hline
	\end{tabular}
	\begin{tablenotes}
        \item[a] $\mathcal{U}$ and $\mathcal{N}$ represent uniform and normal distributions, respectively.
        \item[b] These parameters were fixed in light curve modelling, using the values from \citep{Anderson2014}.
        \item[c] MJD = BJD$_{\textup{TDB}}$ - 2450000.
    \end{tablenotes}
    \end{threeparttable}
\end{table}

\subsection{Spectroscopic light curve}
\label{sec:Spectroscopic light curve} 

To minimize the effect by systematics for the determination of transmission spectrum, similar to other studies~\citep[e.g.,][]{Gibson2013, Carter2020, Chen2021(1)}, we use the common-mode noise derived from the white light curve to correct each spectroscopic light curve. The white-light common-mode noise is the ratio of the white light curve to its best-fit transit model, and then we divided the spectroscopic light curves by this ratio. The corrected spectroscopic light curves were then modelled in a similar manner as introduced in Section~\ref{sec:White light curve}. Each spectroscopic light curve at a given passband was fitted separately to estimate the wavelength-dependent transit parameters $R_{\textup{p}}/R_{\star}$, $u_{1}$ and $u_{2}$. The wavelength independent parameters ($a/R_{\star}$, $i$ and $T_{\textup{mid}}$) are fixed to the median values derived from the white light curve fitting (cf. Table.~\ref{tab:Table 2}). We used the same red noise model as derived from the white light curve fit, but only use time \textit{t} as the additional parameters for Gaussian process inputs.

\begin{figure*}
	\includegraphics[width=\textwidth]{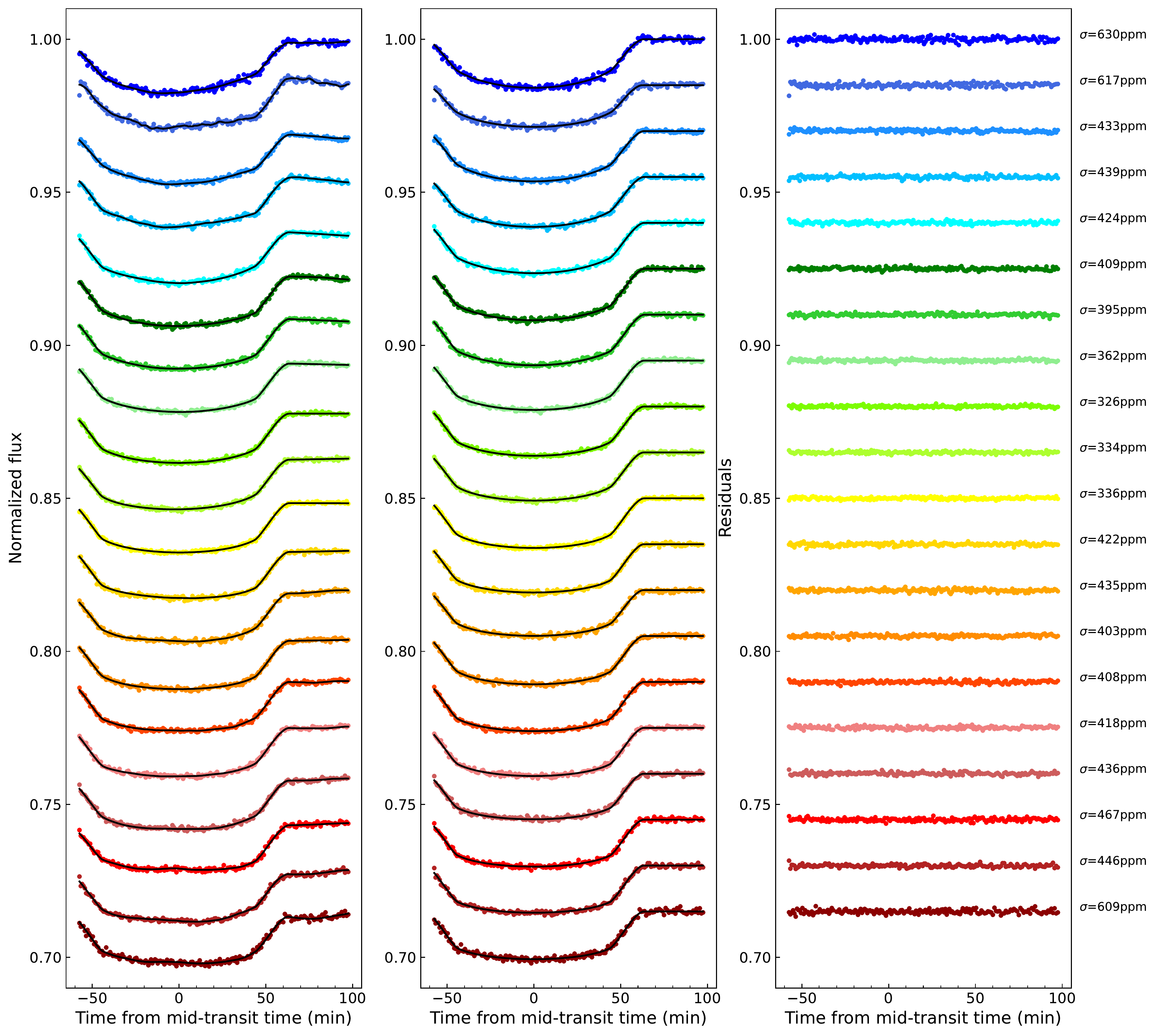}
    \caption{The 20 spectroscopic light curves of Night 1, with each applied a constant vertical offset and different color for distinction. \textit{Left panel}: the raw light curves with common-mode noise removed and the best-fit models (the black solid lines). \textit{Middle panel}: the detrended light curves (colored dot) with the best-fit transit models (the black solid lines). \textit{Right panel}: the observation$-$model residuals.}
    \label{fig:Figure 4}
\end{figure*}

\begin{figure*}
	\includegraphics[width=\textwidth]{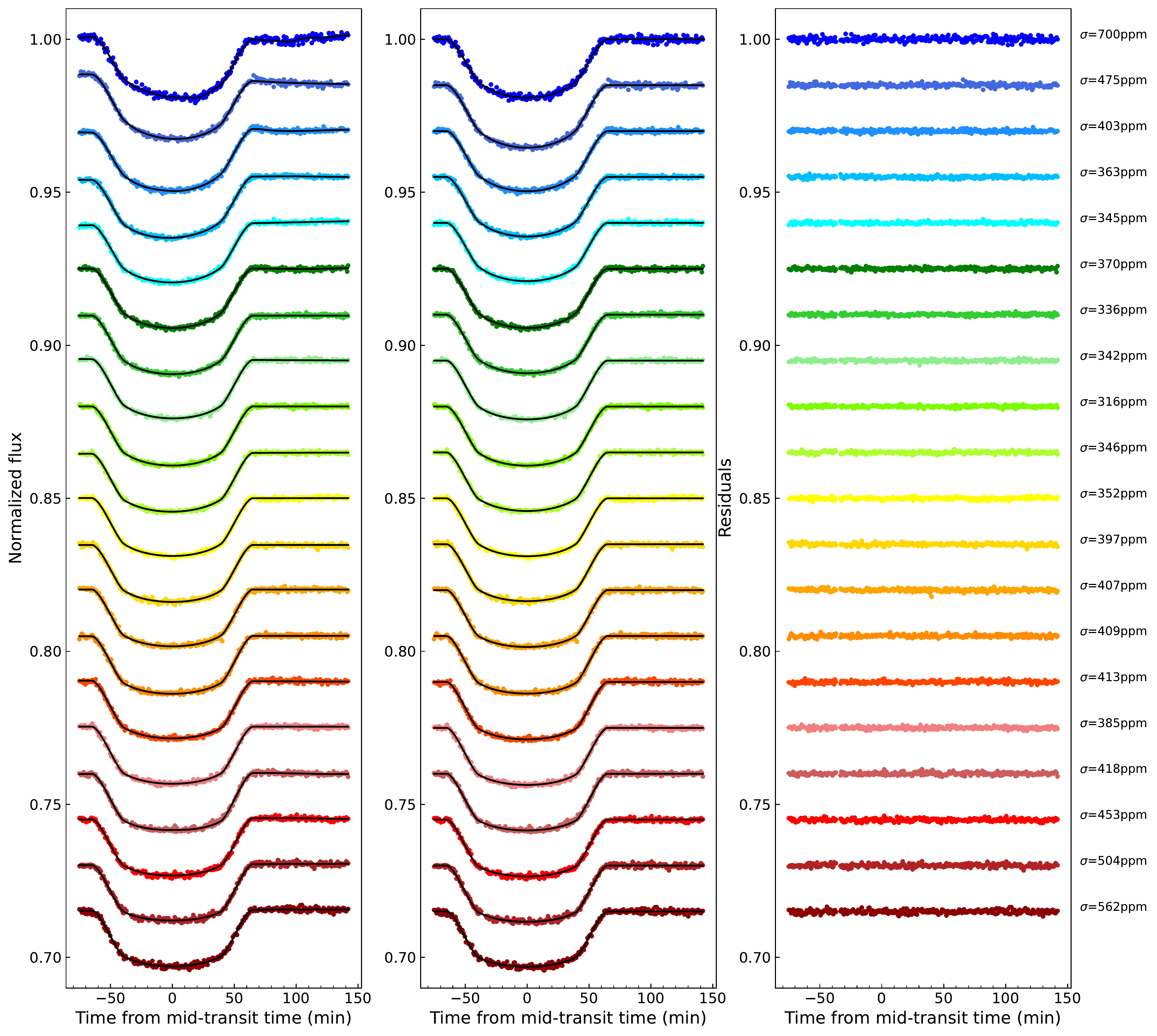}
    \caption{Same as Figure~\ref{fig:Figure 4}, but for the light curves of Night 2.}
    \label{fig:Figure 5}
\end{figure*}

Similarly, the MCMC procedure for each spectral bin consisted of 50 walkers, each with 15000 steps in total, and the first 3000 steps as "burn-in" phase. After all these procedures, we determined the posterior distribution of the parameters of investigation. Figure~\ref{fig:Figure 4} and ~\ref{fig:Figure 5} illustrate the 20 spectroscopic light curves along with the best-fit models for Night 1 \& 2, respectively. The residual standard deviation ($\sigma$) of fitting residuals range between $\sim327$ and $644$\,ppm for Night 1, and range between $\sim328$ and $679$ ppm for Night 2, respectively, or $\sim 1.27-1.79 \times$ and $1.15-1.63 \times$ photon noise correspondingly.

We have tested our data analysis pipeline in deriving spectroscopic light curve by comparing the literature one with the one that is determined by us. For example, we start from the 1D Gemini/GMOS spectra provided by \citet{Wilson2021} (\textit{private communication}) for WASP-121\,b, and obtain the spectroscopic light curve independently using our method as described above. The resulted transmission spectrum is well consistent with the published one in \citet{Wilson2021} (cf. Ouyang et al. \textit{submitted.}) only with a constant offset $\sim$ 472\,ppm.

\section{Transmission spectra and retrieval analysis}
\label{sec:Results}
The planet-to-star radius ratios $R_\textnormal{p}/R_{\star}$ in every spectral channel derived for the two nights are listed in Table~\ref{tab:Table 3} and plotted as transmission spectra in Figure~\ref{fig:Figure 6}. It is shown that the obtained transmission spectra for Night 1 and 2 have different scattering slopes and obvious vertical offsets, with the latter possibly due to the different $R_{\textup{p}}/R_{\star}$ values determined from the white light curve fitting. We also perform a Kolmogorov–Smirnov (K-S) test to verify whether the two transmission spectrum are consistent. The calculated p-value is $5.80\times10^{-10}$, much smaller than 0.05, indicating that the two transmission spectra do not possess similar overall pattern. 

\subsection{The scattering slope}
\label{sec:The slope in transmission spectrum}
\begin{figure}
	\includegraphics[width=\columnwidth]{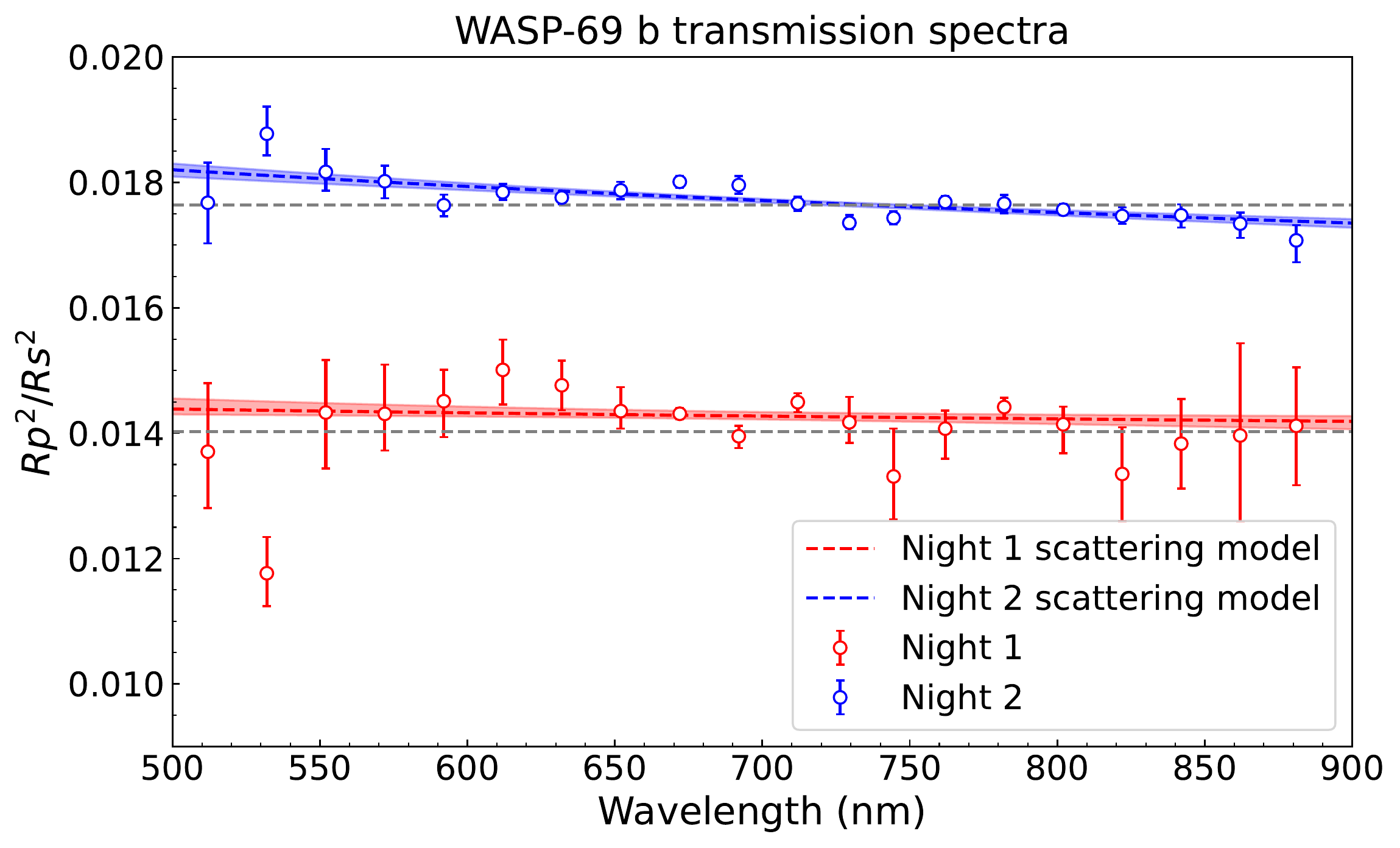}
    \caption{The transmission spectra of WASP-69\,b for Night 1 (the red circles with errorbars) and Night 2 (the blue circles with errorbars). The dashed lines are the white light best fit transit depths. The red and blue dashed lines with colored shaded area indicate the best fit Rayleigh scattering model and the 1$\sigma$ confidence interval, respectively.}
    \label{fig:Figure 6}
\end{figure}

As shown in Figure~\ref{fig:Figure 6}, the two spectra exhibit downward slope towards longer wavelength, which indicates the occurence of Rayleigh scattering occurring in WASP-69\,b's atmosphere. The scattering slope of the planet radius is a function of the wavelength, which is given by \citet{Lecavelier2008}: 
\begin{equation}
\centering
    \frac{\mathrm{d}R_p}{\mathrm{d}\ln{\lambda}} = \alpha H = \alpha\frac{k_B T}{\mu g}
	\label{eq:scattering slope equation}
\end{equation}
where $H$ is the scale height of the planet atmosphere, $T$ is the planetary temperature, $k_\textnormal{B}$ is the Boltzmann constant, $\mu$ is the mean mass of atmospheric particles, and $g$ is the gravity of the planet. The value of $\alpha$ caused purely by Rayleigh scattering is expected to be close to $\alpha \approx -4$ \citep{Lecavelier2008}.

The two spectra have been fitted using linear regression to find the value of $\alpha$, assuming WASP-69\,b has an equilibrium temperature $T_\textup{eq}$ of $963$\,K and log\,$g$ of 2.726 (cgs) \citep[from ][]{Anderson2014}, and it has a hydrogen-dominated atmosphere with the mean mass of atmospheric particles $\mu = 2.37$~\citep{Murgas2020}. The fitting yields $\alpha = -1.25^{+0.90}_{-1.61}$ and $-4.83^{+0.92}_{-0.91}$ for the two spectra respectively (Fig.~\ref{fig:Figure 6}), among which the determined $\alpha$ value of Night 2 is consistent with Rayleigh scattering in hydrogen-dominated atmosphere, but with a slightly smaller $\alpha$ value. Considering the poor weather in Night 1 and the derived transmission spectrum is of large systematics, only the Night 2 transmission spectrum is used for the following analysis. 

The derived $\alpha$ value in this work is marginally consistent with the value of $-3.35\pm0.75$ obtained from their GTC data by \cite{Murgas2020}. We note that the presence of unocculted stellar spots on the stellar disk may introduce similar wavelength dependent variability. Nevertheless, as shown in Section~\ref{sec:Stellar activity}, this is unlikely the case for our data set, although the host star is an active star with evidence of strong emission in Ca~{\sc ii} HK lines with log\,$R_\textnormal{HK}^{\prime}\sim -4.54$~\citep{Anderson2014}. We will discuss the impact of stellar activity on the obtained transmission spectra in Section~\ref{sec:Stellar activity}.

\begin{table}
\renewcommand\arraystretch{1.5}
    \centering
    \caption{The planet-to-star radius ratios of the 20 passbands of Night 1 \& 2.}
    \label{tab:Table 3}
    \begin{threeparttable}
    \resizebox{\columnwidth}{!}{
    \begin{tabular}{cccc} 
        \hline
        Passband (nm) & Center (nm) & $R_\textup{p}/R_{\star}$ (Night 1) & $R_\textup{p}/R_{\star}$ (Night 2)\\
        \hline
        $502-522$ & 512.0 & $0.11707_{-0.00383}^{+0.00467}$ & $0.13294_{-0.00243}^{+0.00241}$ \\
        $522-542$ & 532.0 & $0.10847_{-0.00242}^{+0.00267}$ & $0.13702_{-0.00126}^{+0.00158}$ \\
        $542-562$ & 552.0 & $0.11970_{-0.00373}^{+0.00349}$ & $0.13478_{-0.00112}^{+0.00136}$ \\
        $562-582$ & 572.0 & $0.11962_{-0.00245}^{+0.00328}$ & $0.13422_{-0.00100}^{+0.00093}$ \\
        $582-602$ & 592.0 & $0.12045_{-0.00237}^{+0.00208}$ & $0.13279_{-0.00067}^{+0.00064}$ \\
        $602-622$ & 612.0 & $0.12251_{-0.00225}^{+0.00197}$ & $0.13357_{-0.00045}^{+0.00048}$ \\
        $622-642$ & 632.0 & $0.12151_{-0.00164}^{+0.00162}$ & $0.13325_{-0.00039}^{+0.00038}$ \\
        $642-662$ & 652.0 & $0.11980_{-0.00115}^{+0.00159}$ & $0.13369_{-0.00053}^{+0.00051}$ \\
        $662-682$ & 672.0 & $0.11962_{-0.00038}^{+0.00040}$ & $0.13419_{-0.00036}^{+0.00036}$ \\
        $682-702$ & 692.0 & $0.11812_{-0.00080}^{+0.00070}$ & $0.13400_{-0.00052}^{+0.00054}$ \\
        $702-722$ & 712.0 & $0.12039_{-0.00067}^{+0.00059}$ & $0.13289_{-0.00043}^{+0.00040}$ \\
        $722-737$ & 729.5 & $0.11905_{-0.00138}^{+0.00169}$ & $0.13173_{-0.00039}^{+0.00047}$ \\
        $737-752$ & 744.5 & $0.11537_{-0.00296}^{+0.00330}$ & $0.13203_{-0.00038}^{+0.00041}$ \\
        $752-772$ & 762.0 & $0.11862_{-0.00201}^{+0.00122}$ & $0.13299_{-0.00038}^{+0.00036}$ \\
        $772-792$ & 782.0 & $0.12007_{-0.00075}^{+0.00061}$ & $0.13289_{-0.00057}^{+0.00052}$ \\
        $792-812$ & 802.0 & $0.11892_{-0.00194}^{+0.00119}$ & $0.13253_{-0.00032}^{+0.00033}$ \\
        $812-832$ & 822.0 & $0.11554_{-0.00328}^{+0.00322}$ & $0.13214_{-0.00047}^{+0.00052}$ \\
        $832-852$ & 842.0 & $0.11761_{-0.00304}^{+0.00302}$ & $0.13219_{-0.00073}^{+0.00066}$ \\
        $852-872$ & 862.0 & $0.11816_{-0.00582}^{+0.00623}$ & $0.113168_{-0.00086}^{+0.00067}$ \\
        $892-890$ & 881.0 & $0.11881_{-0.00397}^{+0.00392}$ & $0.13067_{-0.00134}^{+0.00093}$ \\
        \hline
    \end{tabular}
    }
    \end{threeparttable}
\end{table}

\subsection{Retrieval Analysis}
\label{sec:Retrieval Analysis}
We performed spectral retrieval analyses using the \texttt{petitRADTRANS} package \citep{Molliere2019A&A} to generate 1D model transmission spectra, and the \texttt{PyMultiNest}~\citep{Buchner2014} code to calculate the Bayesian evidence $\mathcal{Z}$. \texttt{PyMultiNest} implements the multimodal nested sampling algorithm based on the MultiNest library~\cite{Feroz2009}. As mentioned in \ref{sec:The slope in transmission spectrum}, only the Night 2 SOAR optical transmission spectrum (hereafter the optical data) was used for the retrieval analyses. 

We first performed a retrieval analysis assuming equilibrium chemistry and an isothermal temperature-pressure (T-P) profile represented by the isothermal temperature ($T_\textnormal{iso}$), using two free parameters, i.e., the C/O number ratio and the metallicity [Fe/H]. As described in \citet{Molliere2017}, the mass fraction of each species can be interpolated from a chemical table as a function of $P$, $T$, [Fe/H] and C/O. The list of reactant species includes H$_2$, He, CO, H$_2$O, HCN, C$_2$H$_2$, CH$_4$, PH$_3$, CO$_2$, NH$_3$, H$_2$S, VO, TiO, Na, K, SiO, e$^{-}$, H$^{-}$, H, and FeH. For the retrievals of the optical data, only the absorption of CO$_2$, CO, CH$_4$, TiO, VO, Na, K, and H$_2$O were included. Collision-induced absorption of H$_2$-H$_2$ and H$_2$-He and Rayleigh scattering of H$_2$ and He were also taken into account. The clouds and hazes were parameterized by cloud-top pressure ($P_\textnormal{cloud}$), assuming a full coverage. In addition to the parameters mentioned above, we also set the reference planet radius ($R_\textnormal{p}$) as a free parameter, while the reference pressure $P_0$ at the planet radius is fixed as 0.01\,bar.

However, it is now widely recognized that planet atmospheres are not necessarily in chemical equilibrium\citep{Madhusudhan2011, Molaverdikhani2019, Roudier2021}. Therefore, we also carried out the retrieval analyses assuming free chemistry and an isothermal temperature-pressure profile. The only difference with the equilibrium chemistry model is that the chemical abundance of each species in consideration is allowed to vary freely within its preset boundaries. This approach is in principal more sensitive to individual species, and is used to search for the species that might be responsible for the observed spectral signatures. In this retrieval process, we included the molecules of H$_2$, He, CO$_2$, CO, CH$_4$, TiO, VO, H$_2$O, and atomic Na and K in the full model. We regard H$_2$ and He as filling gases with their abundances fixed as the solar values, while those of the other 8 species to vary freely within our preset limits.

\begin{figure*}
	\includegraphics[width=\textwidth]{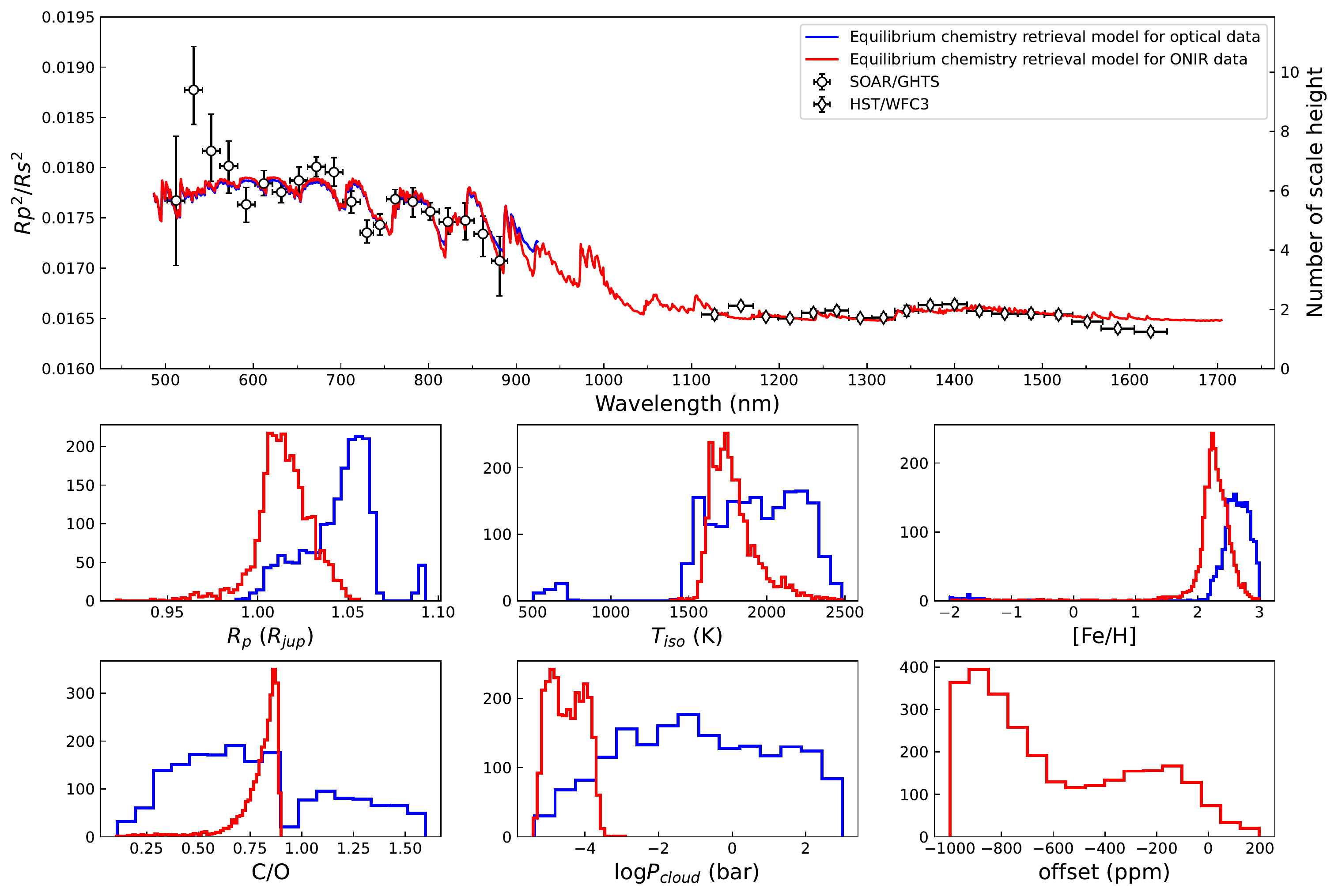}
    \caption{The transmission spectrum of WASP-69\,b and the posterior distributions of the retrieved parameters. First rows: the transmission spectrum obtained by SOAR/GHTS (black open circles) and HST/WFC3 (black open diamond), with the best fit retrieval models assuming equilibrium chemistry. The blue and red solid lines represent the retrieval models for optical data alone and together  with the ONIR data, respectively. Other rows: the posterior distributions of the retrieved parameters for the optical data (blue) and the ONIR data (red), respectively.}
    \label{fig:transmission_spec_eq_model}
\end{figure*}

\begin{figure*}
	\includegraphics[width=\textwidth]{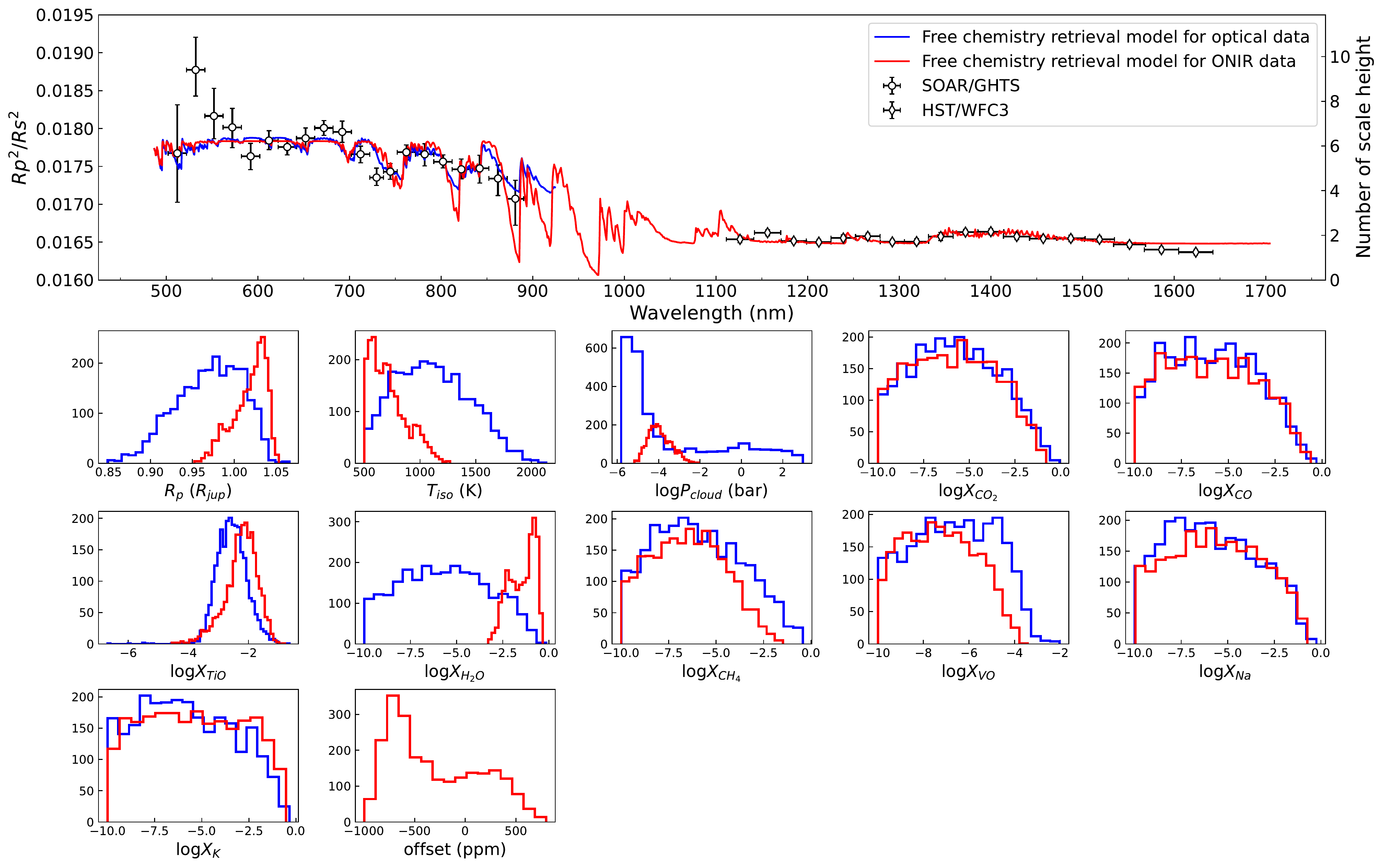}
    \caption{Same as Figure.~\ref{fig:transmission_spec_eq_model}, but retrieval models assuming free chemistry.}
    \label{fig:transmission_spec_free_model}
\end{figure*}

For the optical data alone, the output best-fit transmission spectra and posterior distributions of parameters of investigation assuming equilibrium chemistry and free chemistry are shown in blue in Figures~\ref{fig:transmission_spec_eq_model} \& \ref{fig:transmission_spec_free_model}, respectively. The prior and posterior distributions of all the retrieved parameters for the two retrievals can be found in Table~\ref{tab:retrieval}. The temperatures we obtained for the cases of equilibrium chemistry and free chemistry are $T_\textnormal{iso} = 1937_{-330}^{+302}$ and $1090_{-329}^{+387}$\,K, respectively. We point out a clear detection of TiO in free chemistry retrieval, as show in the its posterior distribution in Figure~\ref{fig:transmission_spec_free_model} suggesting a mass fraction log\,$X_\textup{TiO} = -2.56_{-0.47}^{+0.49}$.

We then perform joint retrieval analysis on the combined data set including our SOAR optical spectra and the HST/WFC3 NIR data from \citet{Tsiaras2018} (hereafter the ONIR data) to achieve a better reliability and to gain additional information. A free parameter is added for the ONIR data retrieval analysis, to account for the offset between the optical and NIR data points. The resulting best-fit models and the posterior distributions of ONIR data are also shown in red in Figures.~\ref{fig:transmission_spec_eq_model} \& \ref{fig:transmission_spec_free_model}. Most parameters retrieved from the ONIR data are in agreement with those from the optical data alone, for both the free and equilibrium chemistry cases, with the former values better constrained. However, we note that the derived $T_\textnormal{iso}$ in ONIR free chemistry retrieval is low ($T_\textnormal{iso} = 689_{-132}^{+213}$), while the deduced H$_2$O and TiO mass ratio are extremely high (log\,$X_\textnormal{H$_{2}$O}$ = $-1.36_{-0.94}^{+0.64}$ and log\,$X_\textnormal{TiO} = -2.18_{-0.57}^{+0.43}$), which we consider implausible.

\begin{figure*}
	\includegraphics[width=\textwidth]{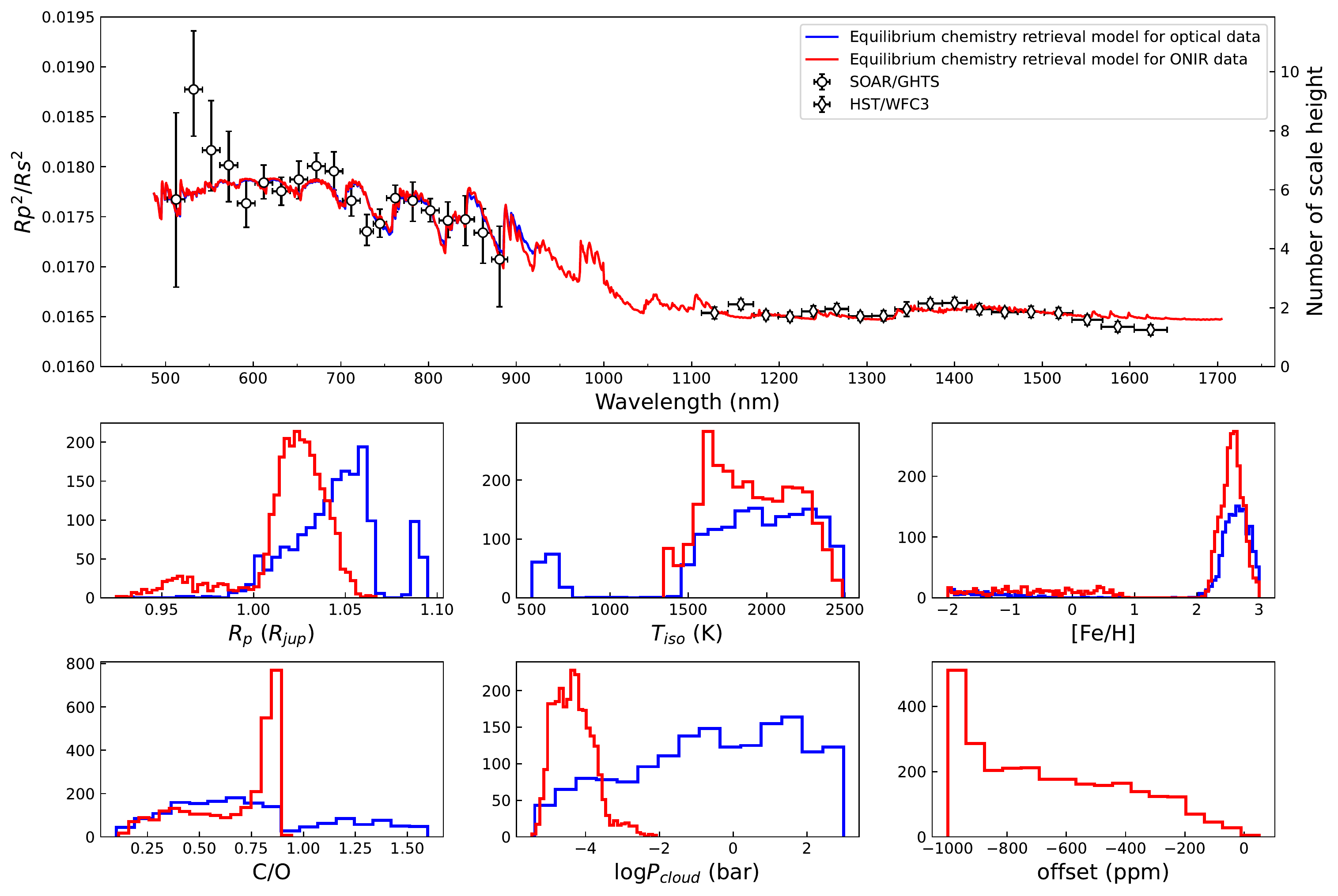}
    \caption{Same as Figure.~\ref{fig:transmission_spec_eq_model}, but after rescaling the uncertainties of transmission spectrum.}
    \label{fig:transmission_spec_eq_model_rescale}
\end{figure*}

\begin{figure*}
	\includegraphics[width=\textwidth]{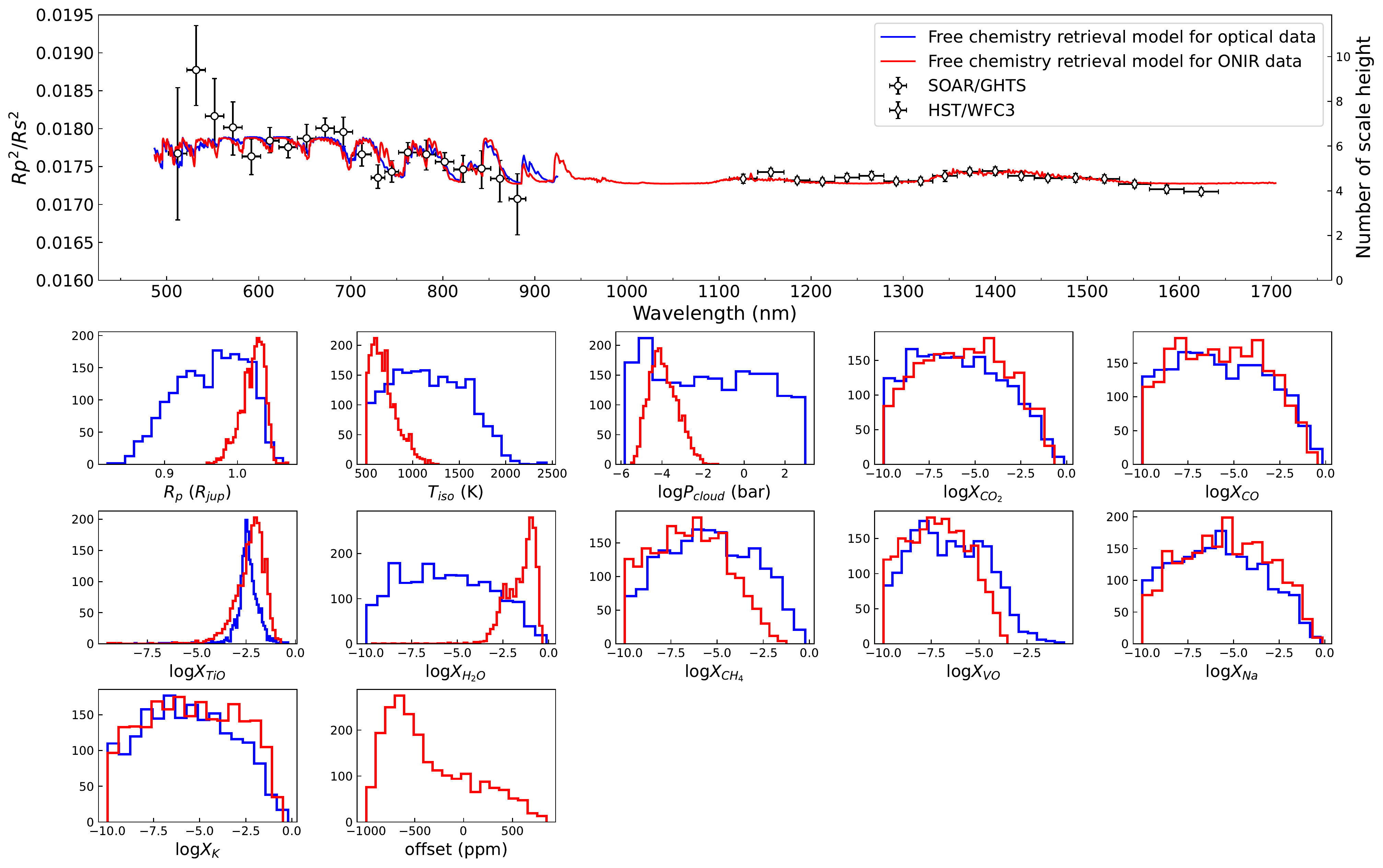}
    \caption{Same as Figure.~\ref{fig:transmission_spec_free_model}, but after rescaling errorbars of transmission spectrum.}
    \label{fig:transmission_spec_free_model_rescale}
\end{figure*}

We note that the $\chi_{\nu}^{2}$ values of all four retrievals are larger than 2, which imply poor fits of the models to the data. One possible reason is that the uncertainties of the obtained transmission spectra are underestimated. We exercise to scale up the errorbars of transmission spectrum to make the best-fit $\chi_{\nu}^{2}$ values close to 1. We first perform a full retrieval for the ONIR data assuming equilibrium chemistry with \texttt{PLATON} package~\citep{Zhang2020}, and determine a scale factor $\eta = 1.357$. Next we use \texttt{petitRADTRANS} to perform independent retrieval analysis on the optical and ONIR transmission spectra with uncertainties rescaled. The best-fit models of the four new retrievals (optical and ONIR data assuming equilibrium and free chemistry) are shown in Figure~\ref{fig:transmission_spec_eq_model_rescale} \& \ref{fig:transmission_spec_free_model_rescale}, and the posterior distributions are listed in Table~\ref{tab:retrieval}. It is shown that the two retrieval results assuming equilibrium chemistry after rescaling seem to be the best ones given the smallest $\chi_{\nu}^{2}$. They are thus served as the bases for the retrieval tests and discussion presented in the next section. We note that the PLATON retrieval also yields a high temperature with $T_\textup{iso} = 1761_{-581}^{+158}$\,K, consistent with the pRT retrieved value (cf. Fig.~\ref{fig:platon_corner}).

\begin{table*}
\renewcommand\arraystretch{1.5}
	\centering
	\caption{The free parameters and the retrieval results for the equilibrium chemistry and free chemistry models.}
	\label{tab:retrieval}
	\begin{threeparttable}
     \resizebox{\textwidth}{!}{
	\begin{tabular}{cccccc} 
		\hline
		Equilibrium chemistry & Prior & \multicolumn{2}{c}{Posterior of optical} & \multicolumn{2}{c}{Posterior of ONIR} \\
        \hline
         &  & before rescaling  & after rescaling  & before rescaling  & after rescaling  \\
		\hline
		$T_\textnormal{iso}$ (K) & $\mathcal{U}(500, 2500)$ & $1937_{-330}^{+302}$ & $1954_{-378}^{+341}$ & $1755_{-103}^{+166}$ & $1877_{-268}^{+344}$ \\
		$R_\textnormal{p}$ ($R_\textnormal{jup}$) & $\mathcal{U}(0.8, 1.2)$ & $1.048_{-0.025}^{+0.011}$ & $1.046_{-0.026}^{+0.015}$ & $1.014_{-0.011}^{+0.014}$ & $1.024_{-0.016}^{+0.014}$ \\
		log\,$P_\textnormal{cloud}$(bar) & $\mathcal{U}(-6, 3)$ & $-1.00_{-2.18}^{+2.53}$ & $-0.41_{-2.77}^{+2.22}$ & $-4.49_{-0.52}^{+0.55}$ & $-4.34_{-0.52}^{+0.57}$ \\
		C/O & $\mathcal{U}(0.1, 1.6)$ & $0.71_{-0.30}^{+0.51}$ & $0.67_{-0.30}^{+0.54}$ & $0.83_{-0.10}^{+0.04}$ & $0.78_{-0.39}^{+0.08}$ \\
		$[\textnormal{Fe/H}]$ & $\mathcal{U}(-2, 3)$ & $2.63_{-0.21}^{+0.21}$ & $2.61_{-0.26}^{+0.22}$ & $2.28_{-0.17}^{+0.20}$ & $2.52_{-2.33}^{+0.20}$ \\
		WFC3 data offset (ppm) & $\mathcal{U}(-1000, 1000)$ & $-$ & $-$ & $-689_{-219}^{+503}$ & $-699_{-249}^{+357}$ \\
		$\chi_{\nu}^{2}$ & - & 2.01 & 1.10 & 2.30 & 1.25 \\
		ln\,$\mathcal{Z}$ & - & 130.4 & 132.5 & 264.5 & 271.4 \\
		\hline
	    Free chemistry & Prior & \multicolumn{2}{c}{Posterior of optical} & \multicolumn{2}{c}{Posterior of ONIR} \\
        \hline
         &  & before rescaling  & after rescaling  & before rescaling  & after rescaling  \\
		\hline
		$T_\textnormal{iso}$ (K) & $\mathcal{U}(500, 2500)$ & $1090_{-329}^{+387}$ & $1148_{-401}^{+461}$ & $689_{-132}^{+213}$ & $669_{-109}^{+170}$ \\
		$R_\textnormal{p}$ ($R_\textnormal{jup}$) & $\mathcal{U}(0.8, 1.2)$ & $0.973_{-0.046}^{+0.037}$ & $0.967_{-0.057}^{+0.047}$ & $1.023_{-0.028}^{+0.015}$ & $1.024_{-0.019}^{+0.013}$ \\
		log\,$P_\textnormal{cloud}$(bar) & $\mathcal{U}(-6, 3)$ & $-4.75_{-0.68}^{+4.72}$ & $-1.73_{-3.00}^{+2.98}$ & $-4.01_{-0.47}^{+0.66}$ & $-3.99_{-0.61}^{+0.78}$ \\
		log\,$X_\textnormal{CO$_2$}$ & $\mathcal{U}(-10, 0)$ & $5.83_{-2.53}^{+2.69}$ & $6.06_{-2.50}^{+2.90}$ & $-5.93_{-2.57}^{+2.60}$ & $5.57_{-2.66}^{+2.55}$ \\
		log\,$X_\textnormal{CO}$ & $\mathcal{U}(-10, 0)$ & $-5.95_{-2.53}^{+2.59}$ & $-5.88_{-2.65}^{+2.99}$ & $-5.95_{-2.60}^{+2.74}$ & $-5.73_{-2.62}^{+2.72}$ \\
		log\,$X_\textnormal{TiO}$ & $\mathcal{U}(-10, 0)$ & $-2.56_{-0.47}^{+0.49}$ & $-2.42_{-0.35}^{+0.48}$ & $-2.18_{-0.57}^{+0.43}$ & $-2.18_{-0.78}^{+0.55}$ \\
		log\,$X_\textnormal{H$_2$O}$ & $\mathcal{U}(-10, 0)$ & $-5.65_{-2.65}^{+2.64}$ & $-5.71_{-2.65}^{+2.99}$ & $-1.36_{-0.94}^{+0.64}$ & $-1.35_{-1.07}^{+0.56}$ \\
		log\,$X_\textnormal{CH$_4$}$ & $\mathcal{U}(-10, 0)$ & $-6.17_{-2.23}^{+2.82}$ & $-5.49_{-2.57}^{+2.73}$ & $-6.50_{-2.14}^{+2.06}$ & $-6.45_{-2.29}^{+2.11}$ \\
		log\,$X_\textnormal{VO}$ & $\mathcal{U}(-10, 0)$ & $-6.67_{-2.04}^{+1.99}$ & $-6.66_{-1.99}^{+2.18}$ & $-7.29_{-1.70}^{+1.81}$ & $-7.18_{-1.75}^{+1.80}$ \\
		log\,$X_\textnormal{Na}$ & $\mathcal{U}(-10, 0)$ & $-6.09_{-2.37}^{+2.96}$ & $-5.88_{-2.57}^{+2.68}$ & $-5.71_{-2.67}^{+2.74}$ & $-5.45_{-2.74}^{+2.65}$ \\
		log\,$X_\textnormal{K}$ & $\mathcal{U}(-10, 0)$ & $-5.95_{-2.58}^{+3.04}$ & $-5.72_{-2.53}^{+2.77}$ & $-5.46_{-2.92}^{+3.10}$ & $-5.43_{-2.81}^{+2.84}$ \\
		WFC3 data offset (ppm) & $\mathcal{U}(-1000, 1000)$ & $-$ &  $-$ & $-413_{-333}^{+673}$ & $-488_{-285}^{+673}$ \\
		$\chi_{\nu}^{2}$ & - & 3.26 & 1.83 & 2.64 & 1.41 \\
		ln\,$\mathcal{Z}$ & - & 127.0 & 129.9 & 267.0 & 272.5 \\
		\hline
	\end{tabular}
    }
    \end{threeparttable}
\end{table*}

\section{Discussion}
\label{sec:Discussion}
\subsection{Comparison with previous work}
\label{sec:Comparison with previous work}
WASP-69\,b is a well-studied hot Jupiter because of its low density and relatively strong atmospheric signals. As introduced in Sect.~\ref{sec:Introduction}, several works have been conducted to study its atmospheric properties using low or high-resolution transmission spectroscopy in recent years. \citet{Tsiaras2018} and \citet{Fisher2018} (hereafter T18 and F18) both presented detailed retrieval analysis of the atmospheres of several exoplanets, including WASP-69\,b. Both studies indicate the presence of non-grey clouds, with a muted water feature. \cite{Murgas2020} (hereafter M20) presented a joint analysis on their GTC/OSIRIS transit data and the HST/WFC3 spectra, and pointed out the existence of Rayleigh scattering in the atmosphere. They also investigated the effect of stellar activity on the transmission spectrum, and concluded that the increase $R_\textnormal{p}/R_{\star}$ ratio towards blue wavelengths could be explained by aerosol rather than unocculted stellar spots or faculae. Then the aerosol-dominated atmosphere with Rayleigh scattering was confirmed by the HST/STIS and HST/WFC3 dataset ~\citep{Estrela2021}, and then by \citet{Khalafinejad2021} (hereafter E21 and K21).

\begin{table*}
\renewcommand\arraystretch{1.5}
	\centering
	\caption{Some retrieval parameters from all the literature low-resolution transmission spectroscopic study of WASP-69\,b.}
	\label{tab:derived_param_comp}
	\begin{threeparttable}
    \resizebox{\textwidth}{!}{
	\begin{tabular}{lcccccc} 
		\hline
		Derived parameters & \citet{Tsiaras2018} & \citet{Fisher2018} & \citet{Murgas2020}$^{a}$ & \citet{Estrela2021} & \citet{Khalafinejad2021} & this work$^{b}$ \\
		\hline
		$T_\textnormal{iso}$ (K) & $493 \pm 153$ & $658_{-107}^{+148}$ & $1227_{-164}^{+133}$ & $> 1450$ & $900 \pm 410$  & $1877_{-268}^{+344}$ \\
        Cloud top pressure (log\,$P_\textnormal{cloud}$(bar))  & $-1.07 \pm 0.99$ & $-$ & $-2.2_{-1.7}^{+1.4}$ & $-1.40_{-0.12}^{+0.10}$ & $-3.1 \pm 1.1$ & $-4.34_{-0.52}^{+0.57}$ \\
        log$_{10}\,Z$ & $-$ & $-$ & $2.7_{-0.2}^{+0.1}$ & $-2.79_{-0.13}^{+0.43}$ & $-$ & $2.52_{-2.33}^{+0.20}$ \\ 
        C/O & $-$ & $-$ & $0.37_{-0.20}^{+0.19}$ & $0.03_{-0.03}^{+1.67}$ & $-$ & $0.78_{-0.39}^{+0.08}$ \\
        log\,$X_\textnormal{H$_2$O}$ & $-3.94 \pm 1.25$ & $-4.24_{-1.09}^{+1.03}$ & $-$ & $-$ & $-2.5 \pm 1.1$ & $-$ \\
		\hline
	\end{tabular}
    }
    \begin{tablenotes}
        \item[a] The retrieval results of their "without stellar spots" assumption.
        \item[b] The retrieval results of after rescaling ONIR data errorbars, assuming equilibrium chemistry.
    \end{tablenotes}
    \end{threeparttable}
\end{table*}

\begin{figure}
	\includegraphics[width=\columnwidth]{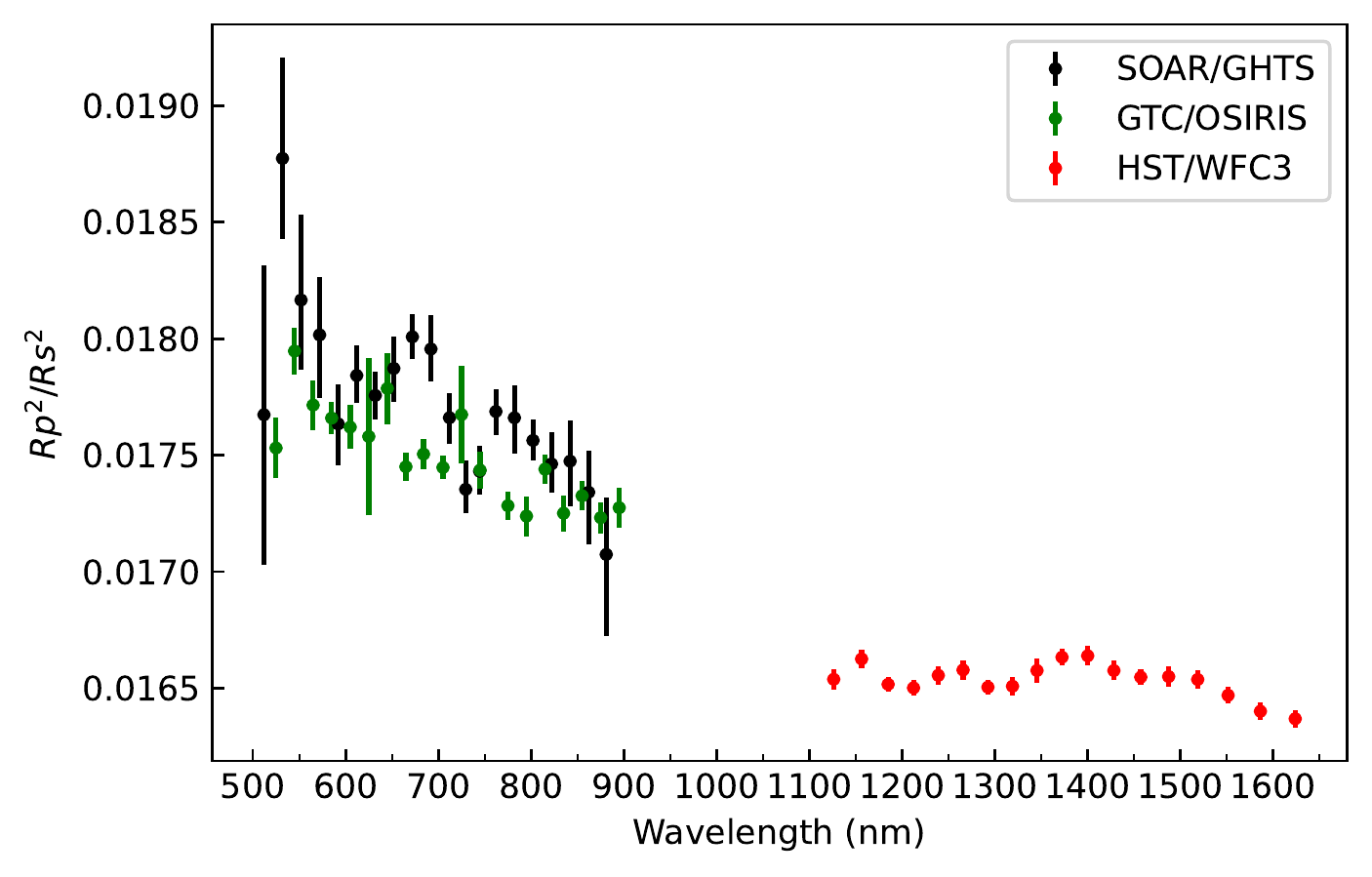}
    \caption{The transmission spectrum of WASP-69\,b obtained by SOAR/GHTS (in black), compared with GTC/OSIRIS (in green) and HST/WFC3 (in red), respectively.}
    \label{fig:transmission_spec_comp}
\end{figure}

As shown in Figure.~\ref{fig:transmission_spec_comp}, our derived optical transmission spectrum is consistent with the GTC/OSIRIS one in the blue and red channels, but is obviously different in the central channels. As expected, our spectrum has larger uncertainties particularly in the most blue and red channels. We have obtained a mean uncertainty $\sim 1.7$ times as large as the GTC/OSIRIS's. In addition, a visual inspection of Fig.~6 in E21 suggests that their G650L spectrum is slightly higher than the M20 one, and is similar to ours.

The scattering slope $\alpha$ ($-4.83^{+0.92}_{-0.91}$) derived from this work is consistent with the value of $\alpha = -3.35 \pm 0.75$ from \citep{Murgas2020} at $\sim1\sigma$ level. The comparison of our results with HST/WFC3 shows a slight offset in $R_\textnormal{p}/R_{\star}$, with a median $\Delta (R_\textnormal{p}/R_{\star})$ of $\sim 0.00728$. This discrepancy could be introduced by the differences between instrument systematics and the settings of orbital parameters including $a/R_{\star}$ and $i$ in the two studies. Such discrepancy have been reported by several earlier works~\citep[e.g., ][]{Alexoudi2018, Murgas2020}. We test the scenario that these two parameters are fixed with $a/R_{\star} = 11.953$ and $i = 86.71^{\circ}$~\citep{Tsiaras2018}, and re-analyze the white light curve and find that the discrepancy still remains, with $\Delta (R_\textnormal{p}/R_{\star}) \sim 0.00695$. The remaining offset is thus likely to be caused by instrumental systematics, which has been taken into consideration in our retrieval analysis. There is also a possibility that the remaining offset is caused by stellar activity, however, this scenario seems to be unpreferred according to our discussion in the following section.

All the atmospheric studies of this planet using low-resolution data mentioned above present their own derived atmosphere parameters, such as metallicities, temperatures and Rayleigh slopes. However, some parameters differ significantly from each other (cf. Table~\ref{tab:derived_param_comp}). For example, although all these works adopted isothermal T-P profile, T18 and F18 deduced $T_\textnormal{iso}$ lower than $T_\textnormal{eq}$ while M20, E21 and this work retrieved higher temperature. On the other hand, we obtain a metallicity about 300 times of solar metallicity, which is well consistent with that from M20 but very different from the E21 value by about 5 magnitudes. As for the C/O ratio, M20 and this work have consistent values, while again E21 does not. We suspect that the large inconsistencies of E21 as compared to the others may arise from their quite different blue slope, while the large slope may be caused physically by the presence of condensates or technically by the treatments of limb darkenning, orbital system parameters assumed (cf. E21).

\subsection{Stellar activity}
\label{sec:Stellar activity}
Stellar activity, such as stellar spots or/and faculae, could contaminate the transmission spectrum in some cases~\citep{Pont2008, Pont2013, McCullough2014}. Several works have confirmed that unocculted spots may cause a blue slope of $R_\textnormal{p}/R_{\star}$ just like Rayleigh scattering~\citep{Rackham2018, Murgas2020, Estrela2021}, while unocculted faculaes may produce a red slope of $R_\textnormal{p}/R_{\star}$~\citep{Rackham2017, Kirk2021, Rathcke2021}. Since WASP-69 is a relatively active K-type star, the impact of its activity needs to be considered for planet transmission spectra. ~\citet{Rackham2018} and references therein adopted an equation to describe the the expected change of apparent radius ratio caused by stellar activity, given by:
\begin{equation}
    \left( \frac{\hat{R}_p}{\hat{R}_s} \right)_{\lambda}^{2} = \left( \frac{R_p}{R_s} \right)^{2} \frac{1}{1-\delta_{\textnormal{spot}}(1-\frac{F_{\lambda}(\textnormal{spot})}{F_{\lambda}(\textnormal{phot})})-\delta_{\textnormal{facu}}(1-\frac{F_{\lambda}(\textnormal{facu})}{F_{\lambda}(\textnormal{phot})})}
	\label{eq:stellar activity}
\end{equation}
where $\delta_{\textnormal{spot}}$ and $\delta_{\textnormal{facu}}$ are the area fraction of the spot and faculae area to the entire stellar disk, while $F_{\lambda}(\textnormal{spot})$, $F_{\lambda}(\textnormal{facu})$ and $F_{\lambda}(\textnormal{phot})$ are the fluxes of stellar spot, faculae and photosphere, respectively.

\citet{Murgas2020} and \citet{Estrela2021} followed to use Eq.~\ref{eq:stellar activity} to verify the scenario that the blue slope seen in the transmission spectrum is totally caused by stellar activity. They both used the \texttt{PLATON} package to retrieve the transmission spectrum, by setting the temperatures and fractions of spots and faculae as free parameters. The best-fit spot and faculae surface coverages and temperatures from \citet{Murgas2020} are $\delta_{\textnormal{spot}} = 0.55_{-0.27}^{+0.30}$, $\delta_{\textnormal{facu}} = 0.15_{-0.13}^{+0.46}$, $T_{\textnormal{spot}} = 4594_{-77}^{+48}$\,K and $T_{\textnormal{facu}} = 4788_{-68}^{+308}$\,K. \citet{Estrela2021} derived the heterogeneity fractional coverage $f_{het} = 0.448 \pm 0.042$ with $T_{\textnormal{spot}} = 4307 \pm 25$\,K and $T_{\textnormal{facu}} = 4604 \pm 24$\,K. Both works suggested a large coverage of stellar spots and faculaes could reproduce their observed slopes of transmission spectrum. However, neither \citet{Murgas2020} nor \citet{Estrela2021} found any evidence of spot-crossing events to support this scenario, based on the follow-up observations made in the discovery paper by \cite{Anderson2014} and their white-light light curves. They both disfavoured the possibility of their observed blue slopes being induced by stellar activity. 

As presented in Fig.~\ref{fig:transmission_spec_comp}, our derived optical transmission spectra is roughly consistent with that from ~\citet{Murgas2020} particularly in the blue and red wavelengths, therefore we should have similar impact from stellar activity. We perform similar retrieval analysis of the stellar contamination for optical data, employing Eq.~\ref{eq:stellar activity}. We consider a simple equilibrium-chemistry atmospheric model assuming that stellar activities have impacts on the blue slope. The retrieved best-fit model and posterior are shown in Fig.~\ref{fig:Figure stellar contam} and Table~\ref{tab:retrieval_stellar_contam}. The derived surface coverages of spots and faculae are $\delta_{\textnormal{spot}} = 0.33_{-0.17}^{+0.27}$ and $\delta_{\textnormal{facu}} = 0.25_{-0.17}^{+0.35}$, and the derived temperatures are $T_{\textnormal{spot}} = 4305_{-297}^{+138}$\,K and $T_{\textnormal{facu}} = 4893_{-105}^{+199}$\,K, respectively, which are slightly different from those obtained from \citet{Murgas2020} and \citet{Estrela2021}. The difference may arise from the $\sim 1.91$ difference in the phase angle of the host star's rotation, given a rotational period $P_{\rm rot}$ of 23 days~\citep{Anderson2014}.

We note that the temperature contrast between the stellar photosphere and the spots is $\Delta T_\textnormal{{phot-spot}} = 410$\,K. Such spot with $\sim$33\% coverage, if crossing, may produce variation of up to $\sim$1100\,ppm in transit light curve, which corresponds to $\sim$6.5\% of transit depth. Such a variation is $\sim2-3$ times of the quoted errors in spectroscopic transit depths, which should be detectable if the spot crossing events occur. As discussed previously, unocculted spots would lead to a slope in the transmission spectrum~\citep{McCullough2014}, and they do actually provide a pretty comparable fit to the data without requiring a very high $T_\textup{iso}$ in our stellar activity retrieval (cf. Table~\ref{tab:retrieval_statistics} \& \ref{tab:retrieval_stellar_contam} and Fig.~\ref{fig:Figure stellar contam}). However, the Bayesian evidence we obtained disfavours this scenario at $\sim 3\sigma$ level. Thus, similar to M20 and E21, we consider stellar activity as a less likely cause for the observed transmission spectrum.
 
\subsection{Tentative detection of TiO} 
\label{sec:Tentative TiO in WASP-69 b atmosphere}

It was proposed and widely accepted for a long while that TiO and VO are the main species responsible for the thermal inversions in the atmospheres of HJs~\citep{Hubeny2003, Fortney2008}. However, only in gaseous state they can induce thermal inversion, thus requires at least the thermal inversion region in the atmosphere to be hot enough ($\sim 1500$\,K for TiO)\citep{Fortney2008}. Therefore, TiO is mostly found in atmosphere of ultra hot Jupiters (UHJs), such as WASP-33\,b~\citep{Haynes2015,Nugroho2017,Cont2021}, WASP-12\,b~\citep{Stevenson2014}, HD\,209458\,b~\citep{Desert2008, Santos2020} and HAT-P-65\,b~\citep{Chen2021(2)}, in which TiO should be mostly in gaseous state. The equilibrium temperature of WASP-69\,b is $T_\textnormal{eq}=963 \pm 18$\,K according to the discovery paper by \cite{Anderson2014}), not high enough the keep TiO in gaseous state. However, recently \citet{Estrela2021} obtained an isothermal temperature $>$1450\,K from their retrieval analysis on their HST STIS and WFC3 data, allowing the presence of TiO gas in WASP-69\,b's atmosphere. They further pointed out the existence of aerosols in the upper atmosphere of WASP-69\,b, and inferred that the photochemical hazes might heat the planet's upper atmosphere, which was later confirmed by \citet{Lavvas2021}. 

As shown in Table~\ref{tab:retrieval}, the retrieved isothermal temperatures from equilibrium chemistry models are higher than 1500\,K, with $\sim 1954$\,K from the optical data and $\sim 1877$\,K from the ONIR data, much higher than the quoted $T_\textnormal{eq}$ from \cite{Anderson2014}. Therefore, for the equilibrium chemistry models, TiO should mostly remain in gaseous state in the atmosphere of WASP-69\,b and should be detectable if abundant. The temperatures retrieved from free chemistry are too low ($\sim 1148$\,K from the optical data and $\sim 669$\,K from ONIR) to sustain abundant gaseous TiO, which we consider to be self-inconsistent and not used as the main data for our following discussion.

\begin{table*}
	\centering
	\caption{Statistics of all retrieval models.}
	\label{tab:retrieval_statistics}
	\begin{threeparttable}
	\begin{tabular}{lccccccccc} 
		\hline
	    $\#$ & Model & \multicolumn{4}{c}{optical} & \multicolumn{4}{c}{ONIR} \\
	    \hline
	     & & dof & $\chi_{\nu}^{2}$ & ln\,$\mathcal{Z}$ & $\Delta$ ln\,$\mathcal{Z}$ & dof & $\chi_{\nu}^{2}$ & ln\,$\mathcal{Z}$ & $\Delta$ ln\,$\mathcal{Z}$ \\
	    \hline
	    \multicolumn{10}{l}{\textit{A. Assuming equilibrium chemistry}} \\
	    1 & Full model & 15 & 1.10 & 132.5 & 0 & 32 & 1.25 & 271.4 & 0 \\
	    2 & No TiO & 15 & 1.32 & 129.6 & -2.9 & 32 & 1.75 & 262.6 & -8.8 \\
	    3 & two-point T-P profile & 12 & 1.27 & 132.9 & 0.4 & 29 & 1.40 & 269.4 & -2.0 \\
		\hline
		\multicolumn{10}{l}{\textit{B. Assuming free chemistry}} \\
		1 & Full model & 9 & 1.83 & 129.9 & 0 & 26 & 1.41 & 272.5 & 0 \\
		2 & No TiO & 10 & 1.74 & 126.7 & -3.2 & 27 & 2.07 & 269.9 & -2.6 \\
		\hline
		\multicolumn{10}{l}{\textit{C. Assuming stellar contamination}} \\
		1 & A1 with stellar contamination & 11 & 1.31 & 129.3 & -3.2 & & & & \\
		\hline
	\end{tabular}
	\end{threeparttable}
\end{table*}

According to our analysis described in Section~\ref{sec:Retrieval Analysis}, the best fit models assuming free chemistry on the optical data and ONIR data both indicate a large fraction of TiO in the atmosphere of WASP-69\,b. The TiO feature look indeed strong with obvious ``zigwig''-shape feature in the optical band ($700-900$\,nm), and the model spectrum matches well with the observed SOAR optical spectrum (cf. Fig.~\ref{fig:transmission_spec_eq_model}). The retrieved TiO mass fraction in the two free chemistry models are $-2.42_{-0.35}^{+0.48}$ and $-2.18_{-0.78}^{+0.55}$\,dex, respectively, for two data sets. Accordingly, the derived TiO mass fraction in the two equilibrium chemistry models on both data sets are $-4.35$ and $-4.21$\,dex, respectively. Therefore, all the models in our consideration imply that TiO is one of the most abundant species in the atmosphere of WASP-69\,b. If confirmed, this will be the first ''classic'' HJ with $T_{\rm eq}\sim 1000$\,K that has been found to possess significant amount of gas TiO. This is particularly interesting but requiring additional confirmation. 

To further verify the presence of TiO, we perform several additional retrieval analyses. Firstly, we run the ``non-TiO'' test by adopting the same input parameters and their priors in both the equilibrium and free chemistry model, as listed in Table~\ref{tab:retrieval}, but only excluding the TiO molecule from consideration. The best fitting model spectra are shown in Figure~\ref{fig:retrieval_notio_model} as the colored solid lines, and the goodness of the fitting are listed in Table~\ref{tab:retrieval_statistics}. It is obvious from Figure~\ref{fig:retrieval_notio_model} that the TiO-excluded model spectra do not match well with the observed spectra, particularly in the optical regime. Table~\ref{tab:retrieval_statistics} confirms that with the best-fit TiO-excluded models do fit the data worse than the best-fit full model, as indicated by the larger values of $\chi_{\nu}^{2}$ and smaller ln\,$\mathcal{Z}$. Therefore, the observed transmission spectra are indeed favouring a large amount of TiO molecules existing in the atmosphere of WASP-69\,b.

Note that for all the models we have investigated above employ isothermal temperature profile, that is, the temperatures for every layer at all altitudes/pressures are the same. This is a simplified but commonly-used T-P profile, which was employed by almost all the previous transmission spectroscopic studies\,\citep[e.g.,][]{Murgas2020,Estrela2021}. To study how more complicated T-P profiles may affect our derived parameters, we follow to employ the widely-used two-point T-P profile from \cite{Brogi2014}, instead of the isothermal temperature profile and performed additional retrieval analysis assuming equilibrium chemistry. This T-P profile assumes isothermal atmosphere at altitudes above the lower pressure point ($T_1$, $P_1$) and below the higher pressure point ($T_2$, $P_2$), and the temperature is assumed to change linearly with log\,$P$ between the two points. The parameters ($T_{1,2}$, $P_{1,2}$) are free parameters to be retrieved. The obtained T-P profiles for the optical and ONIR data are shown in  Fig.~\ref{fig:T_P_profile_2point}, indicating that there is quite some fractional of the atmosphere hotter than 1500\,K. The derived best-fit models have larger $\chi_{\nu}^{2}$ and smaller ln\,$\mathcal{Z}$ compared with those of the full equilibrium chemistry model, as shown in Table~\ref{tab:retrieval_statistics}, suggesting that WASP-69\,b likely possesses an isothermal atmosphere. This result indicates that our assumption of an isothermal atmosphere should be sufficient for our data set.

There is a concern that this pattern may be related to the red noise in the spectroscopic light curves. Therefore, in order to testify this possibility, we run a test by changing the bin size of spectroscopic light curves to be 10 and 30\,nm. We find that the obtained narrower and wider-binned transmission spectra share the very similar trend with the standard spectrum, i.e., with bin size of 20\,nm (cf.Fig.~\ref{fig:tran_spec_binsize_comp}), although the spectroscopic light curve residuals decrease with the bin size increasing. This suggests that the wiggles shown in the SOAR transmission spectrum, i.e., the evidence for TiO, seem not affected by the residual noise.

It is known that TiO absorption may present in the spectra of M stars and late-type K stars (later than K5)~\citep{Morgan1943}, which may imprint on our derived planet transmission spectra. Given our host star is a K5 star, we check carefully in our spectra, but find no evidence for the existence of strong TiO absorption (Fig~\ref{fig:Figure 1}). Mostly importantly, during our data analysis, all the strong features from stars should have been removed clearly, otherwise we should see strong lines from hydrogen and sodium. Another possibility is that TiO may exist in spots on the star which would not be visible in the stellar spectrum of Fig.~\ref{fig:Figure 1} but may imprint TiO as contamination~\citep{Espinoza2019}. However, given the previous tests on the occurrence of star spots, this is quite unlikely the case.

We note that other similar studies on the same target do not find any evidence for the presence of a large amount of TiO gas~\citep{Murgas2020, Estrela2021}, although the latter study has similar TiO-indicative wiggle pattern from 700 to 900\,nm. Actually, their obtained optical transmission spectra are not in good agreement with each other, as discussed in \cite{Estrela2021}. The discrepancy was attributed mostly to the slight differences of the involved parameters used for data analysis and retrieval, and the onset or not of the telluric influence of ground-based or space observations. On the other hand, such discrepancy may also hint for the temporal variability of HJs, as pointed out by \cite{Wilson2021} for WASP-121\,b. Therefore, we find it is likely that the observed wiggles are caused by the presence of gaseous TiO in the WASP-69\,b's atmosphere, and the atmosphere may change over time. However, we could not yet fully exclude the other possibilities including stellar spots and residual red noise until further multi-epoch high-precision observational studies.

\section{Conclusions}
\label{sec:Conclusions}

In this manuscript, we report on the results of two transit observations of the inflated hot Jupiter, WASP-69\,b, obtained by the GHTS at the 4.1\,m SOAR telescope on the nights of July 15 and July 19, 2017. For each night, we obtain a transmission spectrum of WASP-69\,b by determining transit depths in 20 specified spectral bins, but we only employ the Night 2 spectrum for the atmospheric modelling and discussion, as the weather in Night 1 was not good. 

Our derived transmission spectrum shows an increased slope towards blue wavelength. The increased slope may due to the Rayleigh scattering~\citep{Lecavelier2008}. Therefore, we fit the transmission spectrum using a simple model, assuming a hydrogen-dominated atmosphere (the mean mass of atmospheric particles $\mu = 2.37$), and found $\alpha = -4.83^{+0.92}_{-0.91}$. The value is consistent with a purely H-dominated Rayleigh scattering, and agrees marginally with the values obtained from ~\citet{Murgas2020}. The ``zigwig'' shape that is particularly obvious in the $700-900$\,nm regime of the transmission spectrum indicates the presence of TiO in gas state. There is not strong evidence for sodium or potassium. 

We expand the wavelength of the transmission spectrum from optical to near-infrared by adding the HST/WFC3 data from \citealt{Tsiaras2018} into our analysis. Consistent with previous works of WASP-69\,b, we confirm that the observed Rayleigh scattering slope is more likely to be caused by the haze or aerosols rather than stellar activity. Retrieval analysis of the optical data and the ONIR data both point out to a high concentration ($>-4.47$\,dex) of TiO gas, no matter assuming an equilibrium chemistry or not. If confirmed, WASP-69\,b will be the first ``classic'' HJ with TiO detected. Although the equilibrium temperature $T_\textnormal{eq}$ is found to be low~\citep{Anderson2014} that may hamper the presence of gaseous TiO in WASP-69\,b's atmosphere, it was found later that the atmosphere is actually not in equilibrium with derived isothermal temperature significantly higher than $T_{\rm eq}$. \citet{Murgas2020} and \citet{Estrela2021} derived an isothermal temperature of $1227$\,K and $>1450$\,K, respectively, and our obtained temperature is even higher. Worthy to note that \citet{Khalafinejad2021} derived the isothermal temperature consistent with equilibrium temperature independently using the same dataset as \citep{Murgas2020}. A non-TiO test is also performed, resulting that the models with TiO excluded fit worse than those with TiO included as indicated by higher $\chi_{\nu}^{2}$ and smaller ln\,$\mathcal{Z}$ values. The non-detection of TiO in previous works may due to either temporal variability, or the slight differences in data analysis. Further multi-epoch high-precision observations at high resolution using for example HARPS and CARMENES and/or at low-to-medium resolution with JWST are required to solve this discrepancy, and to confirm/reject the hypothesis of the existence of gaseous TiO in WASP-69\.b.

This work is the first attempt to utilize the SOAR telescope for transmission spectroscopy observation. Compared to previous works which usually use large ground-based telescope, our result shows a slightly worse precision and is partially consistent with the transmission spectrum obtained by GTC/OSIRIS or HST/G750L. In general, the mean value of our errorbars is $\sim$ 1.7 times larger than GTC/OSIRIS's, and this discrepancy is reasonable given the difference in collecting area of the two telescopes. We thus conclude that a 4-meter ground-based telescope like SOAR with a stable instrument like GHTS located at a good observing site can be an alternative option in the study of exoplanet atmospheres via transmission spectroscopic observation for at least planets around bright host stars. More studies using GHTS are on their way. 

\section*{Acknowledgements}
We thank the anonymous reviewer for their constructive comments. This research is supported by the National Key R\&D Program of China No.~2019YFA0405102, the National Natural Science Foundation of China grants No. 11988101, 42075123, 42005098, 62127901, the Strategic Priority Research Program of Chinese Academy of Sciences, Grant No.~XDA15072113, the China Manned Space Project with NO. CMS-CSST-2021-B12. Q.-L. O.-Y., M.Z., J.-S.H. are supported by the Chinese Academy of Sciences (CAS), through a grant to the CAS South America Center for Astronomy (CASSACA) in Santiago, Chile.
 
This work is based on observations obtained at the Southern Astrophysical Research (SOAR) telescope, which is a joint project of the Minist\'{e}rio da Ci\^{e}ncia, Tecnologia e Inova\c{c}\~{o}es (MCTI/LNA) do Brasil, the US National Science Foundation’s NOIRLab, the University of North Carolina at Chapel Hill (UNC), and Michigan State University (MSU).
\section*{Data Availability}

The data products and the raw data are available upon request from the author.



\bibliographystyle{mnras}
\bibliography{paper_final} 




\appendix

\section{Additional tables and figures}

\begin{figure*}
	\includegraphics[width=\textwidth]{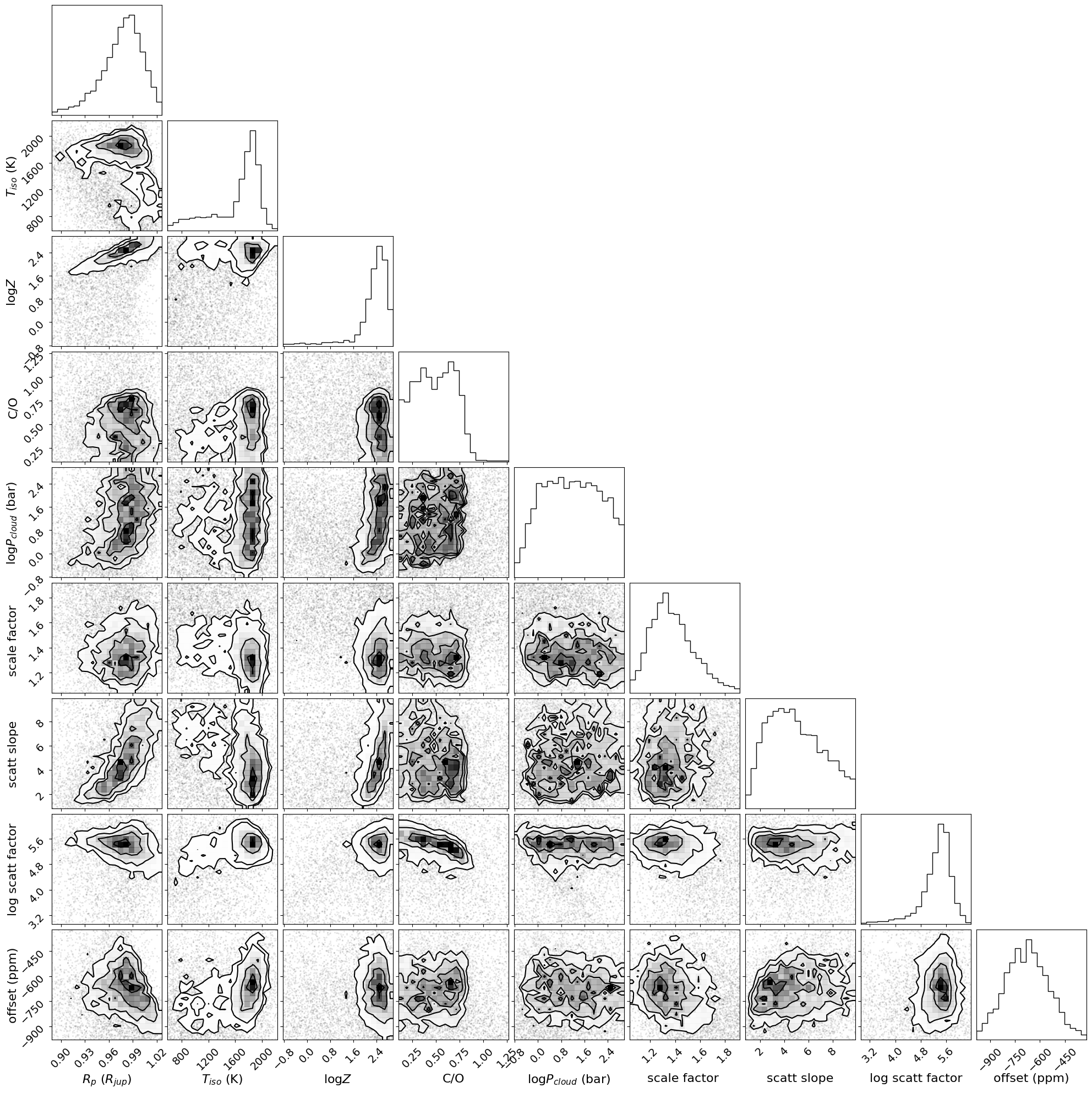}
    \caption{Correlation plot of fitted parameters using \texttt{PLATON} package assuming equilibrium chemistry for ONIR data.}
    \label{fig:platon_corner}
\end{figure*}

\begin{figure*}
	\includegraphics[width=0.95\textwidth]{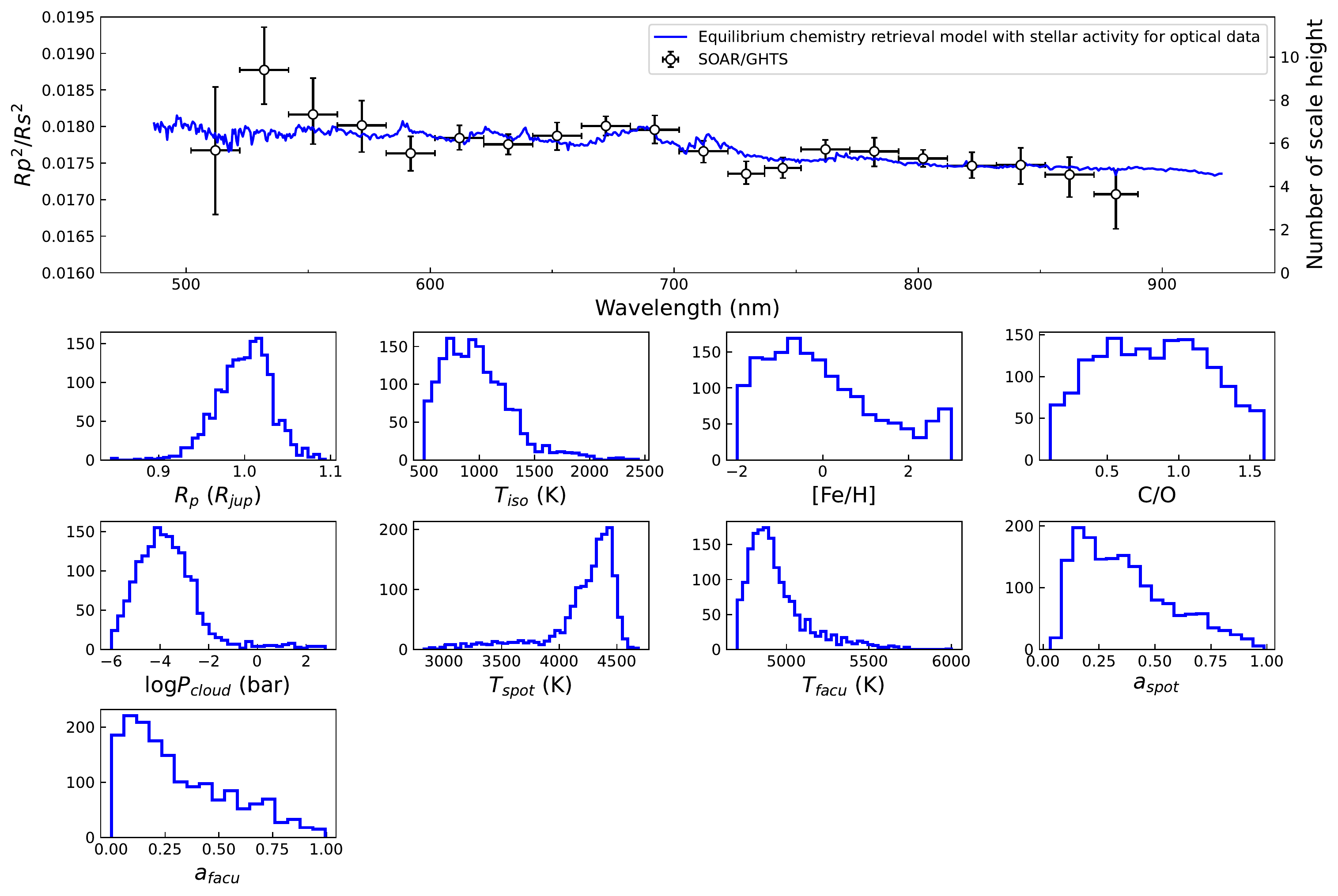}
    \caption{The transmission spectrum of WASP-69\,b and the posterior distributions of retrieved parameters. Top: the transmission spectrum obtained by SOAR/GHTS (black), with the best fit retrieval model assuming equilibrium chemistry with stellar contamination (blue solid line). Middle and bottom: the posterior distributions of retrieved parameters. }
    \label{fig:Figure stellar contam}
\end{figure*}

\begin{table*}
\renewcommand\arraystretch{1.5}
	\centering
	\caption{The same table as Table.~\ref{tab:retrieval}, but for equilibrium chemistry model and considering stellar contamination.}
	\label{tab:retrieval_stellar_contam}
	\begin{threeparttable}
	\begin{tabular}{ccc} 
	    \hline
	    Equilibrium chemistry, stellar contamination & Prior & Posterior of atmosphere with stellar activity\\
		\hline
		$T_\textnormal{iso}$ (K) & $\mathcal{U}(500, 2500)$ & $938_{-260}^{+309}$ \\
		$R_\textnormal{p}$ ($R_\textnormal{jup}$) & $\mathcal{U}(0.8, 1.2)$ & $1.002_{-0.033}^{+0.026}$ \\
		$T_\textnormal{spot}$ (K) & $\mathcal{U}(2700, 4700)$ & $4305_{-297}^{+138}$ \\
		$\delta_\textnormal{spot}$ & $\mathcal{U}(0.0, 1.0)$ & $0.33_{-0.17}^{+0.27}$ \\
		$T_\textnormal{facu}$ (K) & $\mathcal{U}(4700, 7000)$ & $4893_{-105}^{+199}$ \\
		$\delta_\textnormal{facu}$ & $\mathcal{U}(0.0, 1.0)$ & $0.25_{-0.17}^{+0.35}$ \\
		log\,$P_\textnormal{cloud}$ & $\mathcal{U}(-6, 3)$ & $-3.83_{-1.04}^{+1.17}$ \\
		C/O & $\mathcal{U}(0.1, 1.6)$ & $0.84_{-0.44}^{+0.41}$ \\
		$[\textnormal{Fe/H}]$ & $\mathcal{U}(-2, 3)$ & $-0.28_{-1.09}^{+1.72}$ \\
		$\chi_{\nu}^{2}$ & - & 1.31 \\
		ln\,$\mathcal{Z}$ & - & 129.3 \\
		\hline
	\end{tabular}
    \end{threeparttable}
\end{table*}

\begin{figure*}
	\includegraphics[width=0.85\textwidth]{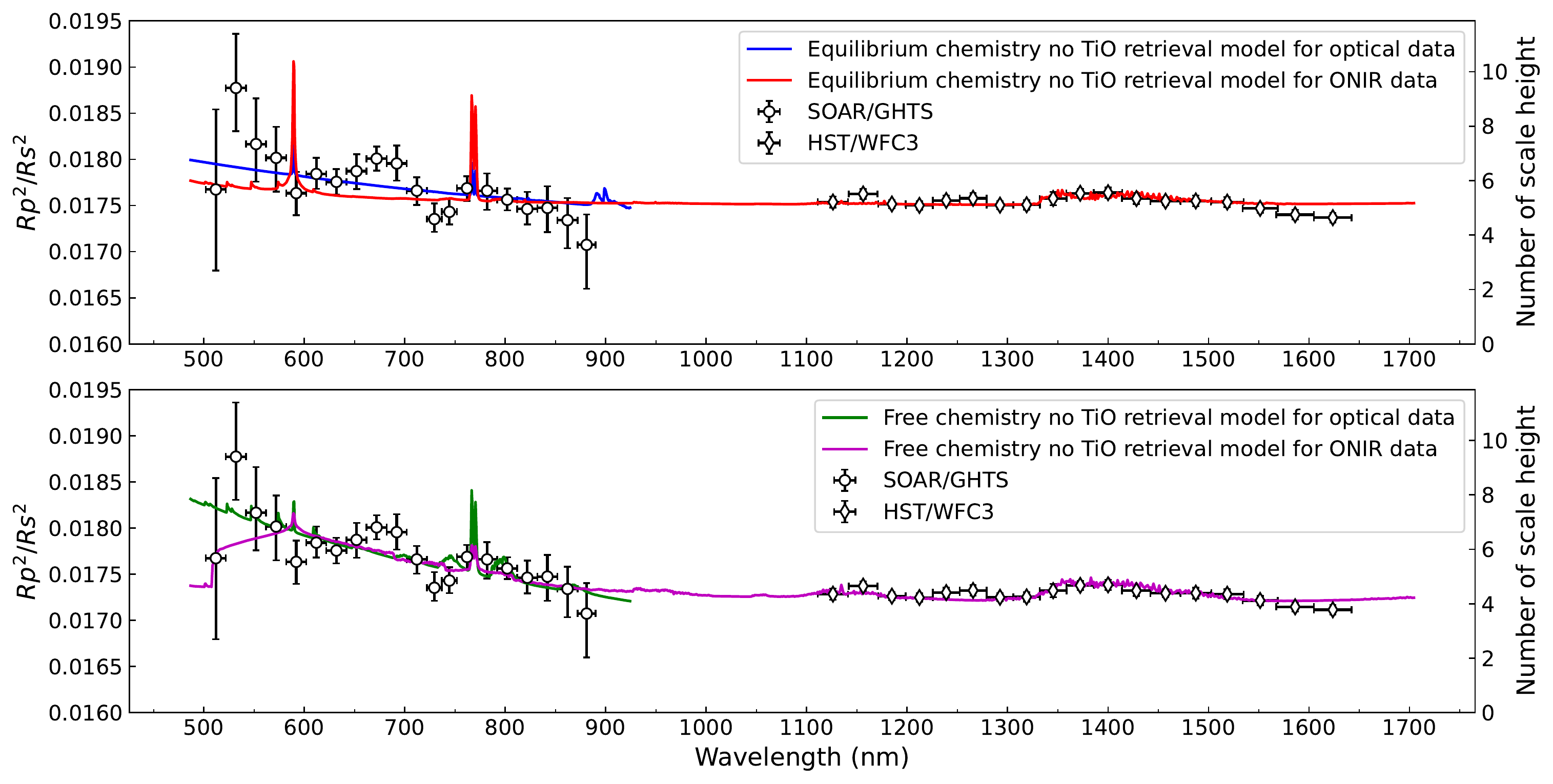}
    \caption{The transmission spectrum of WASP-69\,b with the retrieval models which exclude TiO. First rows: the transmission spectrum obtained by SOAR/GHTS (black hollow circle) and HST/WFC3 (black hollow diamond), with the no TiO retrieval model assuming equilibrium chemistry. The blue solid line is the retrieval model only for optical data, the red solid line is for ONIR data. Second rows: the transmission spectrum with the no TiO retrieval model assuming free chemistry. The green solid line and magenta solid line are the no TiO retrieval model for optical data and ONIR data, respectively.}
    \label{fig:retrieval_notio_model}
\end{figure*}

\begin{figure*}
	\includegraphics[width=0.85\textwidth]{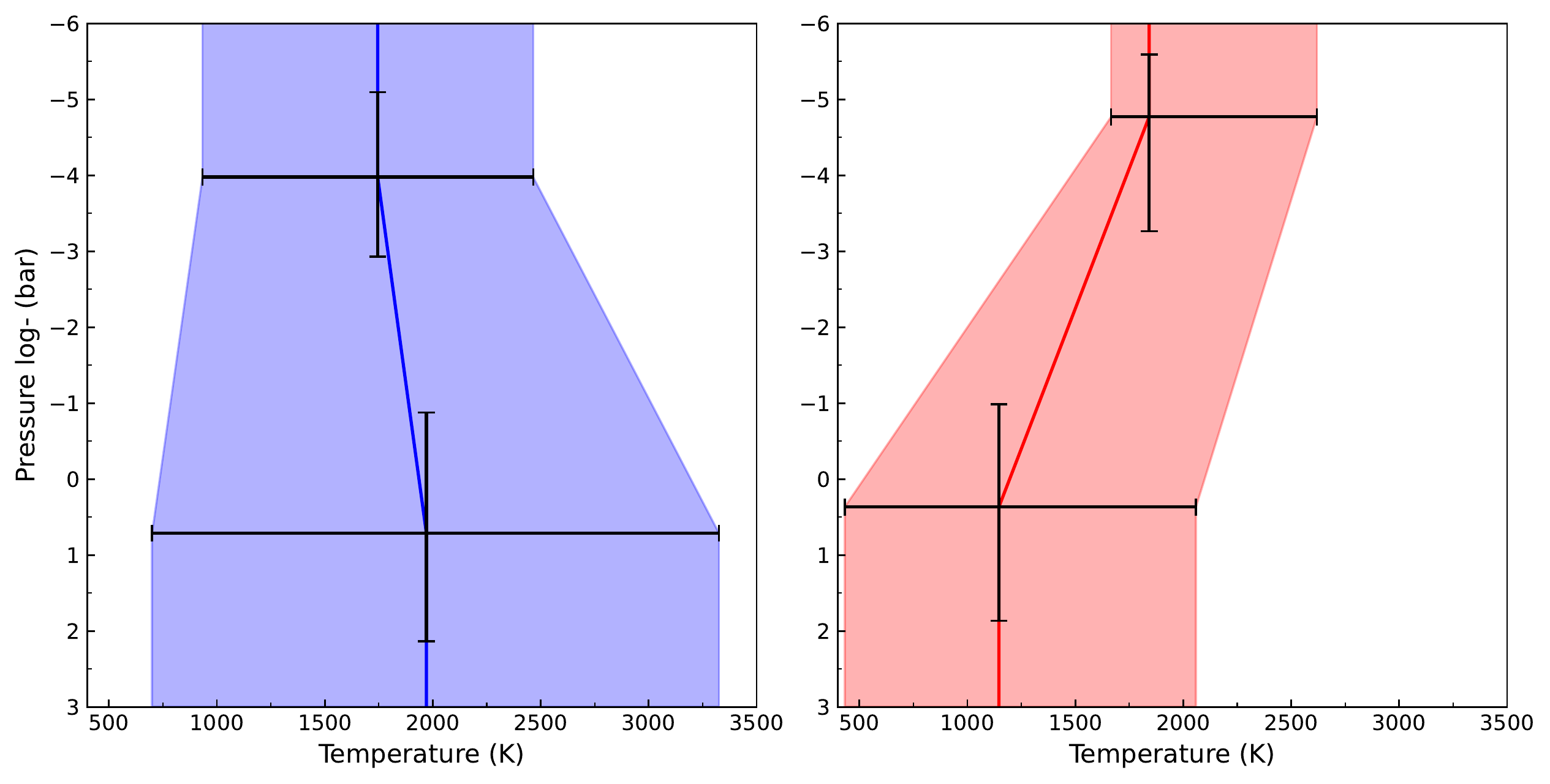}
    \caption{The T-P profile and $1\sigma$ confidence interval retrieved assuming equilibrium chemistry for OPT~ (\textit{Left}) and ONIR~(\textit{Right}), respectively.}
    \label{fig:T_P_profile_2point}
\end{figure*}

\begin{table*}
\renewcommand\arraystretch{1.5}
	\centering
	\caption{The free parameters and their retrieval results for no TiO equilibrium chemistry and free chemistry models.}
	\label{tab:retrieval_notio}
	\begin{threeparttable}
	\begin{tabular}{cccc} 
		\hline
		Equilibrium chemistry (no TiO) & Prior & Posterior of optical & Posterior of ONIR \\
		\hline
		$T_\textnormal{iso}$ (K) & $\mathcal{U}(500, 2500)$ & $642_{-84}^{+68}$ & $1316_{-237}^{+236}$ \\
		$R_\textnormal{p}$ ($R_\textnormal{jup}$) & $\mathcal{U}(0.8, 1.2)$ & $1.090_{-0.003}^{+0.002}$ & $1.086_{-0.025}^{+0.008}$ \\
		log\,$P_\textnormal{cloud}$(bar) & $\mathcal{U}(-6, 3)$ & $1.17_{-1.36}^{+1.23}$ & $-1.48_{-0.98}^{+0.25}$ \\
		C/O & $\mathcal{U}(0.1, 1.6)$ & $0.95_{-0.50}^{+0.40}$ & $0.81_{-0.19}^{+0.07}$ \\
		$[\textnormal{Fe/H}]$ & $\mathcal{U}(-2, 3)$ & $-1.49_{-0.35}^{+0.61}$ & $-1.78_{-0.17}^{+0.55}$ \\
		WFC3 data offset (ppm) & $\mathcal{U}(-1000, 1000)$ & $-$ & $-990_{-7}^{+14}$ \\
		$\chi_{\nu}^{2}$ & $-$ & 1.32 & 1.75 \\
		ln\,$\mathcal{Z}$ & $-$ & 129.6 & 262.6 \\
		\hline
	    Free chemistry (no TiO) & Prior & Posterior of optical & Posterior of ONIR \\
		\hline
		$T_\textnormal{iso}$ (K) & $\mathcal{U}(500, 2500)$ & $913_{-290}^{+478}$ & $628_{-94}^{+238}$ \\
		$R_\textnormal{p}$ ($R_\textnormal{jup}$) & $\mathcal{U}(0.8, 1.2)$ & $1.069_{-0.010}^{+0.036}$ & $1.067_{-0.004}^{+0.003}$ \\
		log\,$P_\textnormal{cloud}$(bar) & $\mathcal{U}(-6, 3)$ & $-0.00_{-2.57}^{+1.79}$ & $-1.70_{-0.59}^{+0.69}$ \\
		log\,$X_\textnormal{CO$_2$}$ & $\mathcal{U}(-10, 0)$ & $-6.00_{-2.56}^{+3.06}$ & $-6.75_{-2.12}^{+2.44}$ \\
		log\,$X_\textnormal{CO}$ & $\mathcal{U}(-10, 0)$ & $-5.75_{-2.68}^{+3.31}$ & $-6.01_{-2.64}^{+2.75}$ \\
		log\,$X_\textnormal{H$_2$O}$ & $\mathcal{U}(-10, 0)$ & $-6.19_{-2.45}^{+2.31}$ & $-3.14_{-0.98}^{+0.73}$ \\
		log\,$X_\textnormal{CH$_4$}$ & $\mathcal{U}(-10, 0)$ & $-6.01_{-2.51}^{+2.47}$ & $-7.49_{-1.65}^{+1.83}$ \\
		log\,$X_\textnormal{VO}$ & $\mathcal{U}(-10, 0)$ & $-9.29_{-0.56}^{+1.62}$ & $-8.43_{-1.01}^{+1.01}$ \\
		log\,$X_\textnormal{Na}$ & $\mathcal{U}(-10, 0)$ & $-3.96_{-2.52}^{+2.44}$ & $-0.20_{-0.12}^{+0.08}$ \\
		log\,$X_\textnormal{K}$ & $\mathcal{U}(-10, 0)$ & $-7.66_{-1.91}^{+3.41}$ & $-5.57_{-2.70}^{+1.75}$ \\
		WFC3 data offset (ppm) & $\mathcal{U}(-1000, 1000)$ & $-$ & $-866_{-89}^{+205}$ \\
		$\chi_{\nu}^{2}$ & $-$ & 1.74 & 2.07 \\
		ln\,$\mathcal{Z}$ & $-$ & 126.7 & 269.9 \\
		\hline
	\end{tabular}
    \end{threeparttable}
\end{table*}

\begin{figure*}
	\includegraphics[width=\textwidth]{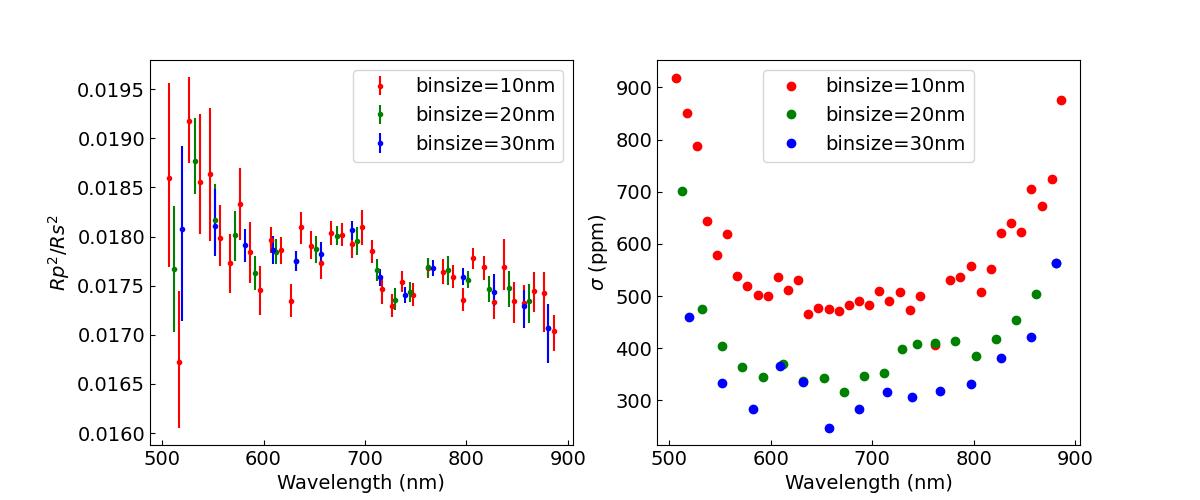}
    \caption{The optical transmission spectrum of WASP-69\,b using different bin size of spectroscopic light curves (shown as different colors), and the corresponding residual standard deviation in different passbands.}
    \label{fig:tran_spec_binsize_comp}
\end{figure*}

\bsp	
\label{lastpage}
\end{document}